\documentclass[twocolumn,aps,showpacs,amssymb,floatfix]{revtex4}
\input epsfig.sty
\newcommand{\be}{\begin{equation}}
\newcommand{\ee}{\end{equation}}
\newcommand{\beq}{\begin{eqnarray}}
\newcommand{\eeq}{\end{eqnarray}}
\usepackage{bm}
\usepackage{graphicx}%

\begin{document}
\setcounter{figure}{\arabic{figure}}

\title{ N to $\Delta$ 
electromagnetic transition form factors from Lattice QCD}
\author{C.~Alexandrou~$^a$,
Ph.\ de Forcrand~$^b$, Th.~Lippert~$^c$, H.~Neff~$^d$,
 J.~W.~Negele~$^e$, K.~Schilling~$^c$, W. Schroers~$^e$ and A. Tsapalis~$^a$}
\affiliation{{$^a$ Department of Physics, University of Cyprus, CY-1678 Nicosia, Cyprus}\\{$^b$ ETH-Z\"urich, CH-8093 Z\"urich and CERN Theory Division, CH-1211 Geneva 23, Switzerland}\\
{$^c$ 
Department of Physics, University of Wuppertal, D-42097 Wuppertal, Germany}\\
{$^d$
Institute of Accelerating Systems and Applications, University of Athens, Athens, Greece and Physics Department, Boston University, U.S.A.}\\
{$^e$ 
Center for Theoretical Physics, Laboratory for
Nuclear Science and Department of Physics, Massachusetts Institute of
Technology, Cambridge, Massachusetts 02139 U.S.A.}}

\date{\today}%

\begin{abstract}
The magnetic dipole, the electric quadrupole and the Coulomb quadrupole
 amplitudes  for the transition $\gamma N\rightarrow \Delta$
are evaluated both in quenched lattice QCD at $\beta=6.0$ and
using two dynamical
Wilson fermions simulated at $\beta=5.6$.
The dipole transition form factor is accurately determined at several
values of  momentum transfer.  
On the lattices studied in this work, the electric quadrupole 
amplitude is found to be non-zero yielding a negative
value for  the ratio, $ R_{EM}$, of electric quadrupole
to magnetic dipole amplitudes  at three values of momentum
transfer.

\end{abstract}

\pacs{11.15.Ha, 12.38.Gc, 12.38.Aw, 12.38.-t, 14.70.Dj}
 
\maketitle
 
 
\section{Introduction}

Recent photoproduction
 experiments on the nucleon at Bates~\cite{Bates} and Jefferson Lab~\cite{Clas}
 have produced accurate  measurements on 
 the ratios of electric and Coulomb quadrupole amplitude to 
the magnetic dipole
 amplitude.
Non-vanishing values for these ratios are thought to be connected
 with nucleon deformation.

Deformation is a common phenomenon in nuclear and atomic physics. 
Classically, the multiphoton coincidence experiment of taking a flash 
photograph or observing an illuminated object distinguishes a deformed 
dumbell from a spherically symmetric sphere. Quantum mechanically, a 
multiphoton coincidence experiment could also determine that a J=0 ground 
state of a diatomic molecule has a deformed shape.  However, usually in 
electromagnetic probes of microscopic systems, we are constrained to make 
measurements associated with one-photon exchange, corresponding to a 
matrix element of a one-body operator. In the case of a diatomic molecule, 
the  one-body  charge density of the J=0 state is spherically symmetric, 
and cannot reveal the deformation that is present in the system.

In many cases, however, when a nuclear or atomic system is well 
approximated by a deformed intrinsic state, it is still possible to 
observe its deformation using a one-body electromagnetic operator. We 
consider here the lowest order electric multipole, the quadrupole moment.  
For an axially deformed object, the quadrupole moment  in the body-fixed 
intrinsic frame is given by
\be
Q_0=\int \> d^3r \rho({\bf r}) \>(3z^2-r^2) \quad
\ee
where $\rho(r)$ is the charge density distribution. If $Q_0$ is positive, 
the object is prolate with the polar axis longer than the equatorial axis.  
In contrast, for an oblate object with the polar axis shorter than the 
equatorial axis, the quadrupole moment is negative.  For collective 
rotation of the deformed intrinsic state~ ~\cite{Bohr}, 
the relation between the 
spectroscopic quadrupole moment, $Q$, measured in the laboratory frame and 
the intrinsic quadrupole moment, $Q_0$, in the body-fixed intrinsic frame 
is given by 
\be
Q=\frac{3K^2-J(J+1)}{(J+1)(2J+3)} Q_0
\ee
where J is the total angular momentum of the system in the lab, K is the 
projection of J onto the z-axis of the body-fixed intrinsic frame, and we 
have considered the sub-state with azimuthal quantum number M=J.  In the 
previous example of the J=0 diatomic molecule, although $Q_0 \ne 0$, Eq. 2 
yields Q=0 so that the deformation, while present, is unobservable. 
Similarly, in the case of a nucleon with J=1/2, Q is zero although $Q_0$ 
may not be. However for the $\Delta$ with  J=3/2, Eq. 2 shows that a 
deformed intrinsic state can be detected by the spectroscopic quadrupole 
moment, Q. The E2 and C2 transition moments between the J=1/2 nucleon and 
J=3/2 nucleon have the same property of revealing the presence of 
deformation in the nucleon, the $\Delta$, or both, and in this work we 
calculate these moments in lattice QCD and provide direct evidence for 
this deformation.

The question whether
the nucleon is deformed from a spherical shape was 
raised twenty years ago~\cite{Isgur} and
it is still unsettled.
On the lattice, hadron wave functions obtained via density-density correlators
can  provide information on the deformation of particles of spin higher than 1/2~\cite{wfs}. 
This approach yields no information on the deformation of the nucleon for the same reason as 
the vanishing of its spectroscopic 
quadrupole moment. 
This is why in lattice studies, like in experiment,
one looks for quadrupole strength
in the  $\gamma N \> \rightarrow \> \Delta$ transition to extract information on
 the nucleon deformation.

State-of-the-art lattice QCD calculations can yield model independent 
results on these matrix elements and provide direct comparison with experiment.
Spin-parity selection rules allow a magnetic dipole, M1, an electric
quadrupole, E2, and a Coulomb quadrupole, C2, amplitude.
If both the nucleon and the $\Delta$
are spherical, then E2 and C2 are
expected to be zero. Although M1 is indeed the dominant amplitude,
there is mounting experimental evidence over  a range of momentum transfer
that E2 and C2 are
 non-zero~\cite{Bates,Clas}. A recent analysis of experimental results
on  the values of E2/M1 and  C2/M1 
is shown to be incompatible with a spherical nucleon~\cite{cnp}.

 Understanding
 the origin of a non-zero deformation is an important theoretical
issue, which depends on QCD dynamics.
The physical origin of non-zero  E2 and C2 amplitudes is
attributed to different mechanisms in the various models:
In the constituent non-relativistic quark model the 
deformation was originally   explained 
by  the color-magnetic hyperfine inter-quark potential arising from  one gluon
exchange and producing 
 a $D$-state admixture in  the singlet-quark
wave functions of the  nucleon and the $\Delta$~\cite{Isgur}.
The deformation due to the hyperfine interaction was also studied in
'relativized' quark models. In both cases the deformation obtained,
measured by the ratio  E2/M1,
is smaller than that found experimentally.

In the context
of the constituent quark model, it was recently proposed 
that elimination of gluonic and quark-antiquark
pairs leads to two-body contributions in the charge and vector 
current operators
that produce a non-zero quadrupole moment in agreement
with experiment using only s-wave
 functions for the nucleon and the $\Delta$~\cite{Buchmann}.  
In  cloudy non-relativistic  models~\cite{Buchmann} 
as well as in chiral or cloudy bag models~\cite{chiral} 
the deformation arises because
of the asymmetric pion cloud whereas in soliton
models it is thought to be due to the  non-linear pion field interactions. 
Most of the results obtained in 
 cloudy baryon models predict 
values smaller than the experimentally measured
ones suggesting that the deformation can not be entirely 
attributed  to the pion cloud. 
This is agreement with lattice results~\cite{wfs}
where a non-zero deformation is observed in the case of the rho meson even in
the quenched approximation where
pion cloud contributions are omitted.

In the present work we will 
 compare
 quenched and unquenched results for the transition
matrix element $\gamma N\rightarrow \Delta$ in order  to  
examine sea quark  contributions to the deformation~\cite{All}.
An early, pioneering 
 lattice QCD study~\cite{Leinweber}
with a limited number of
quenched configurations yielded an inconclusive result for the ratio
of the electric quadrupole to magnetic dipole amplitudes, referred to as
EMR or $R_{EM}$,
  since
a zero value could not be statistically excluded. However the theoretical 
framework that it provided is still applicable 
and we will apply similar techniques to the present study making
a number of improvements:
(i) We use smearing techniques, which very effectively filter
the ground state so that the time independent physical 
observables can be extracted
from the correlators. (ii) The quenched calculation is done on
two volumes at the same parameters to check the volume dependence
of the results.
(iii) Using a large quenched lattice allows us to simulate smaller quark masses,
the lightest giving a ratio of pion  to rho mass of 50 percent.
(iv) For each lattice momentum transfer
${\bf q}$ we calculate the multipoles 
both in the rest frame of the nucleon and of the $\Delta$. These
two different choices of kinematics enable the evaluation of the transition
form factors
at two different 
four-momentum
 transfers.
(v) We study 
dynamical quark effects by evaluating the form factors 
using  the SESAM configurations~\cite{SESAM}
that were produced at $\beta=5.6$ 
on a lattice of size $16^3\times 32$ using Hybrid Monte Carlo
for two degenerate flavours of dynamical Wilson fermions.
vi) In all cases
we use more configurations to improve the statistics.

\section{Lattice matrix elements}

\begin{figure}[h]
\epsfxsize=7.5truecm
\epsfysize=5.5truecm
\mbox{\epsfbox{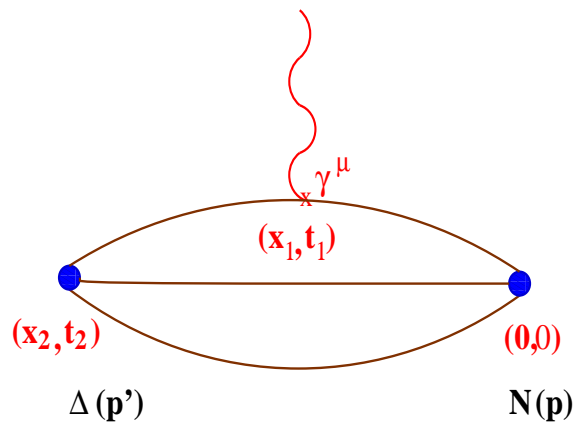}}
\caption{$N\gamma \rightarrow \Delta$ matrix element. The photon couples
to one of  the quarks in the nucleon
  at a fixed time $t_1$ to produce a $\Delta$.} 
\label{fig:diagram}
\end{figure}

The current matrix element for the $\gamma N \> \rightarrow \> \Delta$
transition with on-shell nucleon and $\Delta$ states and real or
virtual photons is shown schematically in Fig.~\ref{fig:diagram}.
It has the form~\cite{Jones73}

\beq
 \langle \; \Delta (p',s') \; | j_\mu | \; N (p,s) \rangle &=& \nonumber \\  
&\>& \hspace*{-4cm} i   \sqrt{\frac{2}{3}} \biggl(\frac{m_{\Delta}\; m_N}{E_{\Delta}({\bf p}^\prime)\;E_N({\bf p})}\biggr)^{1/2} 
  \bar{u}_\tau (p',s') {\cal O}^{\tau \mu} u(p,s) \;
\label{DjN}
\eeq
where $p(s)$ and $p'(s')$ denote initial and final momenta (spins) and 
$ u_\tau (p',s')$ is a spin-vector in the Rarita-Schwinger formalism.

The operator 
${\cal O}^{\tau \mu}$  can be decomposed in terms of the Sachs form factors
as
\be
{\cal O}^{\tau \mu} =
  {\cal G}_{M1}(q^2) K^{\tau \mu}_{M1} 
+{\cal G}_{E2}(q^2) K^{\tau \mu}_{E2} 
+{\cal G}_{C2}(q^2) K^{\tau \mu}_{C2} \;,
\ee
where the magnetic dipole, ${\cal G}_{M1}$, the electric quadrupole, 
${\cal G}_{E2}$,
 and the Coulomb 
quadrupole, ${\cal G}_{C2}$, form factors depend on the momentum
transfer $q^2 = (p'-p)^2$. The kinematical functions 
$K^{\tau \mu}$ depend on $p,\>p',\>M_N$ and $M_\Delta$ and their expressions
are given in ref.~\cite{Jones73}.
The reason for 
using this parametrization for a lattice computation, 
as pointed out in ref.~\cite{Leinweber}, is that
the Sachs form factors  do not depend strongly on the difference
between the 
nucleon and the $\Delta$ mass. 
From the Sachs form factors one can evaluate 
 the ratio 
$E2/M1$ referred to above as 
EMR or $R_{EM}$,
and  the analogous ratio of the Coulomb quadrupole amplitude
 to the magnetic dipole amplitude $C2/M1$, known as CMR or $R_{SM}$,
which are the 
target of recent experiments.
Using the relations given in refs.~\cite{Jones73,Gellas}
the ratios $R_{EM}$ and   $R_{SM}$
in
 the  rest frame of the $\Delta$  
are obtained from the Sachs form factors via 
\be
 R_{EM}= -\frac{{\cal G}_{E2}(q^2)}{{\cal G}_{M1}(q^2)} 
\label{EMR}
\ee
and
\be R_{SM}=-\frac{|{\bf q}|}{2m_\Delta}\;\frac{{\cal G}_{C2}(q^2)}{{\cal G}_{M1}(q^2)} \quad.
\label{CMR}
\ee

The lattice construction of the appropriate matrix elements 
for the evaluation of these form factors follows
closely that of ref.~\cite{Leinweber}.
The  computationally most demanding part in this evaluation is the
calculation
 of the 
 three-point correlation function that corresponds to the diagram shown
in Fig.~\ref{fig:diagram} and it is given by 
\begin{widetext}
\be
\langle G^{\Delta j^\mu N}_{\sigma} 
(t_2, t_1 ; {\bf p}^{\;\prime}, {\bf p}; \Gamma) \rangle =
\sum_{{\bf x}_2, \;{\bf x}_1}
\exp(-i {\bf p}^{\;\prime} \cdot {\bf x}_2 )  
\exp(+i ({\bf p}^{\;\prime} -{\bf p}) \cdot {\bf x}_1 ) \;  
\Gamma^{\beta \alpha}
\langle \;\Omega \; | \; T\left[\chi^{\alpha}_{\sigma}({\bf x}_2,t_2) 
j^{\mu}({\bf x}_1,t_1) \bar{\chi}^{\beta} ({\bf 0},0) \right]
\; | \;\Omega \;\rangle   \; ,
\label{Delta N}
\ee
where an initial state with the quantum numbers of the  
nucleon is created 
at time zero and the final state with the quantum numbers of the
$\Delta$ is annihilated at a later time $t_2$. The photon couples
to one of the quarks in the nucleon at an intermediate time $t_1$
producing a $\Delta$.
The corresponding three-point correlation function for the transition 
$\Delta \rightarrow \gamma N$ is given by

\be 
\langle G^{N j^\mu \Delta}_{\sigma} 
(t_2, t_1 ; {\bf p}^{\;\prime}, {\bf p}; \Gamma) \rangle =
\sum_{{\bf x}_2, \;{\bf x}_1}
\exp(-i {\bf p}^{\;\prime} \cdot {\bf x}_2 )  
\exp(+i ({\bf p}^{\;\prime} -{\bf p}) \cdot {\bf x}_1 ) \; \Gamma^{\beta \alpha}
\langle \;\Omega \; | \; T\left[\chi^{\alpha}({\bf x}_2,t_2) 
j^{\mu}({\bf x}_1,t_1) \bar{\chi}^{\beta}_{\sigma} ({\bf 0},0) \right]
\; | \;\Omega \;\rangle   \quad. 
\label{N Delta}
\ee
For the spin-$\frac{1}{2}$ source ,~$\chi ^p ({\bf x},t)$, and the 
spin-$\frac{3}{2}$ source, 
$\chi ^{\Delta^{+}}_\sigma  ({\bf x},t)$, we use the interpolating fields 
\be
\chi ^p (x) = \epsilon^{a b c}\; \left[ u^{T\; a}(x)\; C \gamma_5
d^b(x) \right]\; u^c(x)
\ee
\be
\chi ^{\Delta^{+}}_\sigma  (x) = \frac{1}{\sqrt{3}} \epsilon^{a b c} \Big
\lbrace
2 \left[ u^{T a}(x)\; C \gamma_\sigma d^b(x) \right]u^c(x) \;
+\; \left[ u^{T a}(x)\; C \gamma_\sigma u^b(x) \right]d^c(x) \Big \rbrace
\ee
and for the projection matrices for the Dirac indices
\be
\Gamma_i = \frac{1}{2}
\left(\begin{array}{cc} \sigma_i & 0 \\ 0 & 0 \end{array}
\right) \;\;, \;\;\;\;
\Gamma_4 = \frac{1}{2}
\left(\begin{array}{cc} I & 0 \\ 0 & 0 \end{array}
\right) \;\; .
\ee

For large Euclidean time separations 
$t_2 -t_1 \gg 1$ and $t_1 \gg 1$,
the time dependence and field normalization constants are 
canceled
 in the following ratio~\cite{Leinweber}

\beq
R_\sigma (t_2, t_1; {\bf p}^{\; \prime}, {\bf p}\; ; \Gamma ; \mu) &=&\large{
 \left [\frac{
\langle G^{\Delta j^\mu N}_{\sigma} (t_2, t_1 ; {\bf p}^{\;\prime}, {\bf p};
\Gamma ) \rangle \;
\langle G^{N j^\mu \Delta}_{\sigma} (t_2, t_1 ; -{\bf p}, -{\bf p}^{\;\prime};
\Gamma^\dagger ) \rangle }
{
\langle \delta_{ij} G^{\Delta \Delta}_{ij}(t_2,{\bf p}^{\; \prime};
\Gamma_4) \rangle \;
\langle G^{NN} (t_2, -{\bf p} ; \Gamma_4) \rangle } \right]^{1/2}
} \nonumber \\
&\;&\stackrel{t_2 -t_1 \gg 1, t_1 \gg 1}{\Rightarrow}
\Pi_{\sigma}({\bf p}^{\; \prime}, {\bf p}\; ; \Gamma ; \mu) \; ,
\label{R-ratio}
\eeq
where $ G^{NN}$ and $ G^{\Delta \Delta}_{ij}$ are the 
 nucleon and $\Delta$ two point functions given  respectively by
\beq
\langle G^{NN} (t, {\bf p} ; \Gamma) \rangle &=&  \sum_{{\bf x}}
e^{-i {\bf p} \cdot {\bf x} } \; \Gamma^{\beta \alpha}\;\langle \Omega |\;T\;\chi^{\alpha}({\bf x},t) 
 \bar{\chi}^{\beta} ({\bf 0},0)  
\; |  \Omega\;\rangle \nonumber \\
\langle G^{\Delta \Delta}_{\sigma\tau} (t, {\bf p} ; \Gamma) \rangle &=&  \sum_{{\bf x}}
e^{-i {\bf p} \cdot {\bf x} } \; \Gamma^{\beta \alpha}\;\langle \Omega |\;T\;\chi^{\alpha}_{\sigma}({\bf x},t) 
 \bar{\chi}^{\beta}_{\tau} ({\bf 0},0)  
\; | \Omega \;\rangle \quad. \nonumber \\
&\>&
\label{NN}
\eeq
The phase in Eq.~\ref{R-ratio} is the same as that of
$G^{\Delta j^\mu N}_{\sigma} (t_2, t_1 ; {\bf p}^{\;\prime}, {\bf p})$
since formally  we have
\be
\Pi_{\sigma}({\bf p}^{\; \prime}, {\bf p}\; ; \Gamma ; \mu)
= \biggl(\frac{E_\Delta + m_\Delta}{E_\Delta}\biggr)^{-1/2}\;
 \biggl(1+\frac{{\bf q}^2}{3m_\Delta^2}\biggl)^{-1/2} \;
\biggl(\frac{E_N + m_N}{2 E_N}\biggr)^{-1/2} \;
 \frac{\langle G^{\Delta j^\mu N}_{\sigma} (t_2, t_1 ; {\bf p}^{\;\prime}, {\bf p};\;\Gamma)\rangle}{Z_N \; Z_\Delta e^{-E_\Delta(t_2-t_1)} \; e^{-E_N t_1}}
\ee

We use the lattice conserved   electromagnetic current,   $j^\mu (x)$,
given by

\be
j^\mu (x) = \sum_{f} Q_{f} \kappa_{f} \lbrace
\bar{\psi}^{f} (x + \hat{\mu})(1 + \gamma_\mu)
U^{\mu \dagger} (x) \psi^{f} (x) 
-\bar{\psi}^{f} (x)(1 - \gamma_\mu)
U^{\mu} (x) \psi^{f} (x + \hat{\mu}) \rbrace
\ee
symmetrized on site $x$ by taking
$
j^\mu (x) \rightarrow \left[ j^\mu (x) + j^\mu (x - \hat \mu) \right]/ 2
,$ where $Q_f$ is the charge of a quark of flavour $f$ and $\kappa_f$
is its hopping parameter.

In the nucleon
laboratory frame  ${\bf p} = 0$ and
$ {\bf p}^{\;\prime}={\bf q}$.
The Sachs form factors can be extracted
from the  plateau values of $ \Pi_{\sigma}({\bf p}^{\; \prime}, {\bf p}\; ; \Gamma ;
\mu)$ for specific combinations of matrices $\Gamma$ and $\Delta$ 
vector indices
$\sigma$. The expressions for general momentum transfer $\bf{q}$
are obtained 
using the standard Euclidean non-relativistic representation for the 
$\gamma$ matrices~\cite{Munster} with $\epsilon^{1234}=1$. The kinematical
functions $K^{\tau \mu}$  in Euclidean space are given by
\beq
 K^{\tau \mu}_{M1} & = & -\frac{3}{(m_\Delta+m_N)^2+Q^2 } \; 
\frac{(m_\Delta+m_N)}{2 m_N}\;
i\epsilon^{\tau \mu \alpha \beta} p^{\alpha}p^{\prime\; \beta} \nonumber \\
  K^{\tau \mu}_{E2} & = & -  K^{\tau \mu}_{M1} + 6\Omega^{-1}(Q^2)\; 
\frac{(m_\Delta+m_N)}{2 m_N} \;
2i\gamma_5 \;\epsilon^{\tau \lambda \alpha \beta} p^{\alpha}p^{\prime\;\beta} 
\epsilon^{\mu \lambda \gamma \delta} p^{\gamma}p^{\prime\;\delta} 
\nonumber \\
K^{\tau \mu}_{C2} & = &-6\Omega^{-1}(Q^2)\; \frac{(m_\Delta+m_N)}{2 m_N} \;
i\gamma_5 \; Q^{\tau} \left(Q^2 P^{\mu} - Q.P Q^{\mu}\right)
\eeq
with $\Omega(Q^2) =  \left[(m_\Delta+m_N)^2+Q^2\right]\left [(m_\Delta-m_N)^2+Q^2\right]$ and ${\bf Q}={\bf q}$, $Q^4=iq^0$ is the lattice
 momentum transfer giving $Q^2=-q^2$. By $p^{\alpha}$ and $p^{\prime \> \beta}$
we now denote Euclidean space momenta defined analogously to $Q^{\mu}$.
The Rarita-Schwinger spin sum for the $\Delta$  in Euclidean space is given by
\be
\sum_s u_\sigma(p,s)\bar{u}_\tau(p,s) = \frac{-i\gamma . p+m_\Delta}{2m_\Delta} \left[\delta_{\sigma\;\tau} +\frac{2p_\sigma p_\tau}{3m_\Delta^2}
-i\frac{p_\sigma\gamma_{\tau}-p_\tau\gamma_{\sigma}}{3m_\Delta}
-\frac{1}{3}\gamma_\sigma \gamma_\tau \right]
\ee 
and the Dirac spin sum 
\be
\sum_s u(p,s)\bar{u}(p,s) =\frac{-i\gamma .p+m_N}{2m_N} \quad.
\ee

We generalize in what follows the expressions of ref.~\cite{Leinweber}
to allow momentum transfers in any spatial direction. 
By selecting the time component of the current the  ${\cal G}_{C2}$ form factor is
extracted from

\beq
{\cal{G}}_{C2}& =& C({\bf q}^2)\;\biggl[\frac{2\;m_\Delta}{\delta^{lk}{\bf q}^2-q^kq^l(1+2E_{\Delta}/m_{\Delta})}\biggr] \;  \Pi_l ({\bf q}, 0\; ; i\Gamma_k ; 4)  \nonumber \\
C({\bf q}^2)&=& \sqrt{\frac{3}{2}} \; \frac{4 E_\Delta m_N}{m_N+m_\Delta}\;
\biggl(\frac{E_\Delta + m_\Delta}{E_\Delta}\biggr)^{1/2}\;
 \biggl(1+\frac{{\bf q}^2}{3m_\Delta ^2}\biggl)^{1/2}  
 \; ,
\label{C2}
\eeq
where $E_\Delta=\sqrt{{\bf p}^{\prime \; 2}+m_\Delta^2}$ and the indices 
$k$ and $l$ denote spatial directions.
By selecting the spatial component of the current 
the ${\cal G}_{M1}$ and ${\cal G}_{E2}$ form factors  are extracted  from

\beq
{\cal{G}}_{M1}& =& C({\bf q}^2) \epsilon^{\sigma l k 4} \frac{1}{q^k} \Pi_\sigma ({\bf q}, 0\; ; i\Gamma_4 ; l) \nonumber \\
&=& C({\bf q}^2) \frac{1}{(q^k)^2-(q^l)^2}\biggl[q^k\Pi_l ({\bf q}, 0\; ; \Gamma_k ; l) - q^l\Pi_k ({\bf q}, 0\; ; \Gamma_l ; k) 
- \frac{m_\Delta}{E_\Delta}\biggl({q^k}\Pi_k ({\bf q}, 0\; ; \Gamma_l ; l) - {q^l}\Pi_l ({\bf q}, 0\; ; \Gamma_k ; k) \biggr) \biggr] \nonumber \\
&\>&
\label{M1}
\eeq
and 
\be
{\cal{G}}_{E2} = \frac{C({\bf q}^2)}{3}\;
\frac{1}{(q^k)^2-(q^l)^2}\biggl[{q^k}\Pi_l ({\bf q}, 0\; ; \Gamma_k ; l) - {q^l}\Pi_k ({\bf q}, 0\; ; \Gamma_l ; k) 
+ \frac{m_\Delta}{E_\Delta}\biggl({q^k}\Pi_k ({\bf q}, 0\; ; \Gamma_l ; l) - {q^l}\Pi_l ({\bf q}, 0\; ; \Gamma_k ; k) \biggr) \biggr] 
\label{E2}
\ee
provided $q^k\ne q^l$.
If  we consider a momentum transfer  that has  zero component along 
the current direction Eqs.~\ref{M1} and \ref{E2} simplify to
\be
\left (
\begin{array}{c} {\cal G}_{M1}\\
                3{\cal G}_{E2} \end{array} \right )  
= \frac{C({\bf q}^2)}{q^k} \biggl[\Pi_l ({\bf q}, 0\; ; \Gamma_k ; l)
\stackrel{-}{+}
       \frac{m_\Delta}{E_\Delta} \Pi_k ({\bf q}, 0\; ; \Gamma_l ; l)\biggr]
\label{M1;E2}
\ee

Another possibility is  to extract the transition form factors by using,
instead of $R_{\sigma}$, 
the ratio
\be
R^{(1)}_{\sigma}(t_2,t_1;{\bf p}^{\; \prime}, {\bf p}\; ; \Gamma ; \mu)
= \frac{\langle G^{\Delta j^\mu N}_{\sigma} (t_2, t_1 ; {\bf p}^{\;\prime}, {\bf p};\;\Gamma)\rangle}{\langle \delta_{ij} G^{\Delta \Delta}_{ij}(t_2,{\bf p}^{\; \prime};
\Gamma_4) \rangle \;\Biggl[\langle G^{NN} (2t_1, {\bf p} ; \Gamma_4)\rangle\; / 
\; \langle \delta_{ij} G^{\Delta \Delta}_{ij}(2t_1,{\bf p}^{\;\prime};
\Gamma_4) \rangle \Biggl]^{1/2} }
\label{R1-ratio}
\ee
or equivalently
\be
R^{(2)}_{\sigma}(t_2,t_1;{\bf p}^{\; \prime}, {\bf p}\; ; \Gamma ; \mu)
= \frac{\langle G^{N j^\mu \Delta}_{\sigma} (t_2, t_1 ; {\bf p}^{\;\prime}, {\bf p};\;\Gamma)\rangle}{\langle  G^{NN}(t_2,{\bf p}^{\; \prime};
\Gamma_4) \rangle \;\Biggl[\langle\delta_{ij}  G^{\Delta\Delta}_{ij} (2t_1, {\bf p} ; \Gamma_4)\rangle\; / 
\; \langle  G^{NN}(2t_1,{\bf p}^{\; \prime};\Gamma_4) \rangle \Biggl]^{1/2} }
\label{R2-ratio}
\ee

\end{widetext}

In this work we 
choose 
the current along the z-direction
and consider 
momentum  transfers
 along the x-axis.
In particular we  consider the lowest allowed lattice momentum transfer
${\bf q}=  (2\pi/Na,0,0)$,
where $a$ is the lattice spacing and $N$ the spatial
lattice size, with the exception of the
large quenched lattice where we also consider momentum transfer 
${\bf q}=  (4\pi/Na,0,0)$.
For our choice of the momentum ${\bf q}$ it is reasonable to take 
in Eqs.~\ref{R-ratio}, \ref{R1-ratio} and \ref{R2-ratio}
 the more symmetric combination
$3/2(G^{\Delta\Delta}_{22}+G^{\Delta\Delta}_{33})$
instead of $\delta_{ij} G^{\Delta\Delta}_{ij}$. 
With this replacement  the second square root involving ${\bf q}^2$ in 
the overall factor $C({\bf q}^2)$ given in  Eq.~\ref{C2} is absent.
Using the fact that the momentum transfer has a component only in the
x-direction Eqs.~\ref{C2} and \ref{M1;E2} simplify to 

\be
{\cal{G}}_{C2}^{(a)} = C_0\;\frac{m_{\Delta}}{{\bf q}^2}  \;  \frac{m_\Delta}{E_\Delta} \;\Pi_1 ({\bf q}, 0\; ; -i\Gamma_1 ; 4)  
\label{C2 11}
\ee
\beq
{\cal{G}}_{C2}^{(b)}& =& C_0 \;\frac{2m_{\Delta}}{{\bf q}^2}   \; \Pi_2 ({\bf q}, 0\; ; i\Gamma_2 ; 4)   \nonumber \\
 &=& 
C_0 \;\frac{2m_{\Delta}}{{\bf q}^2} \; \Pi_3 ({\bf q}, 0\; ; i\Gamma_3 ; 4)
\label{C2 22+33}
\eeq
and
\be
{\cal{G}}_{M1}^{(a)} =   C_0 \; \frac{1}{|{\bf q}|}\; \Pi_2 ({\bf q}, 0\; ; i\Gamma_4 ; 3)  
\label{M1 24}
\ee
\be
{\cal{G}}_{M1}^{(b)} =    C_0 \; \frac{1}{|{\bf q}|}\;\biggl[\Pi_3 ({\bf q}, 0\; ; \Gamma_1 ; 3)
- \frac{m_\Delta}{E_\Delta}\Pi_1 ({\bf q}, 0\; ; \Gamma_3 ; 3) \biggr] 
\label{M1 31;13}
\ee
\be
{\cal{G}}_{E2}^{(a)} =  \frac{C_0}{3}\;\frac{1}{|{\bf q}|} \; \biggl[2\;\Pi_3 ({\bf q}, 0\; ; \Gamma_1 ; 3) -\Pi_2 ({\bf q}, 0\; ; i\Gamma_4 ; 3)\biggr] \\
\label{E2 24;31}
\ee
\be
{\cal{G}}_{E2}^{(b)}  =\frac{C_0}{3}\;\frac{1}{|{\bf q}|} \;
 \biggl[\Pi_3 ({\bf q}, 0\; ; \Gamma_1 ; 3)
+ \frac{m_\Delta}{E_\Delta}\Pi_1 ({\bf q}, 0\; ; \Gamma_3 ; 3) \biggr]
\label{E2 31;13}
\ee
where $C_0$ is obtained from $C({\bf q}^2)$ given in Eq.~\ref{C2}
by omitting the second square root~\cite{error}.
When the $\Delta$ is produced at rest,  the factors
$m_\Delta/E_\Delta$ in Eqs.~\ref{C2 11}, \ref{M1 31;13} and \ref{E2 31;13}
 are absent and 
$C_0 \rightarrow \sqrt{\frac{3}{2}}\;\frac{4m_N E_N}{m_\Delta+m_N}\;\sqrt{\frac{E_N+m_N}{E_N}}$.
The resulting formulas
then agree with those given
in ref.~\cite{Leinweber}.

\begin{figure}[h]
\epsfxsize=8.5truecm
\epsfysize=6.5truecm
\mbox{\epsfbox{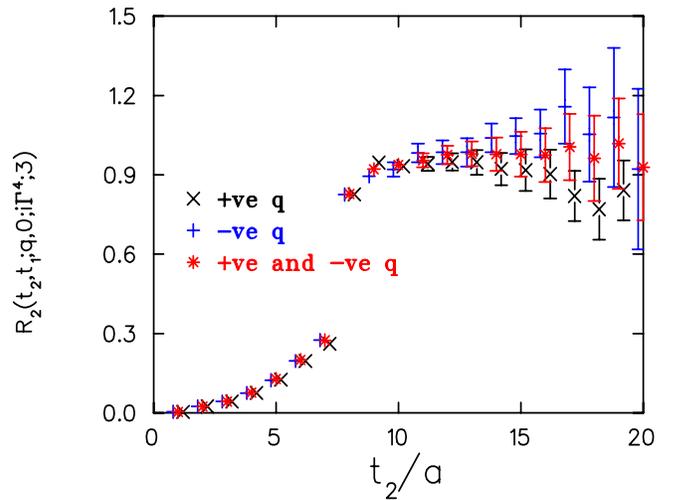}}
\caption{The ratio 
$R_2 (t_2,t_1;{\bf q}, 0; i\Gamma_4 ; 3)$.
Data  with a photon carrying 
   momentum ${\bf q}$ {\it and} $-{\bf q}$  
averaged over 50 quenched configurations are denoted
 by the stars. Data with a photon  carrying either 
positive or negative ${\bf q}$
averaged over 100 configurations
are shown by the x-symbols  and the crosses respectively.
These results are for a lattice of size $32^3\times 64$
at $\kappa=0.1558$. The current couples to the quark at $t_1/a=8$.}
\label{fig:noisereduction}
\end{figure}

\begin{figure}[h]
\epsfxsize=7.0truecm
\epsfysize=11truecm
\mbox{\epsfbox{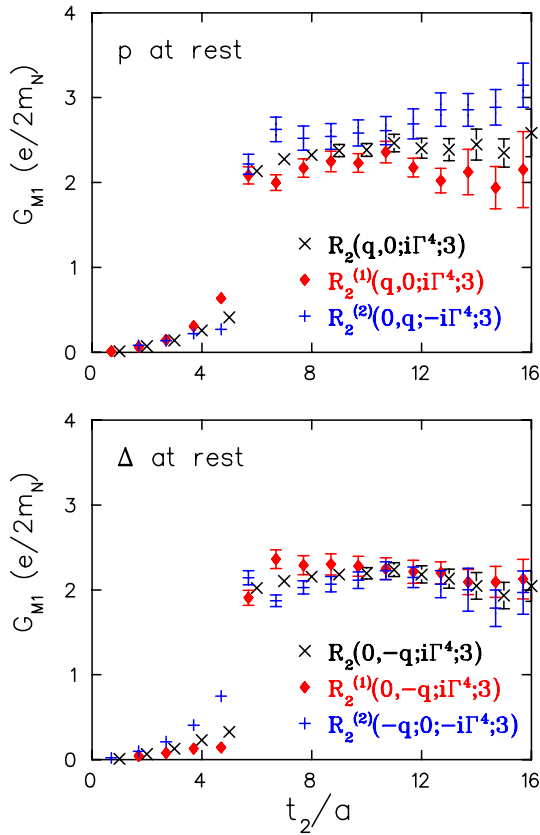}}  
\caption{${\cal G}_{M1}^{(a)}$ in units of natural magnetons
 using the ratios $R_\sigma$ (x's),  $R_\sigma^{(1)}$ (diamonds) 
and  $R_\sigma^{(2)}$ (crosses)
 with the nucleon at rest (top)  and with the  $\Delta$ at rest (bottom)
for dynamical quarks at $\kappa=0.1565$. 
The current couples to the quark at $t_1/a=6$.}
\label{fig:G vs R}
\end{figure}

\begin{figure}[h]
\epsfxsize=7.0truecm
\epsfysize=11truecm
\mbox{\epsfbox{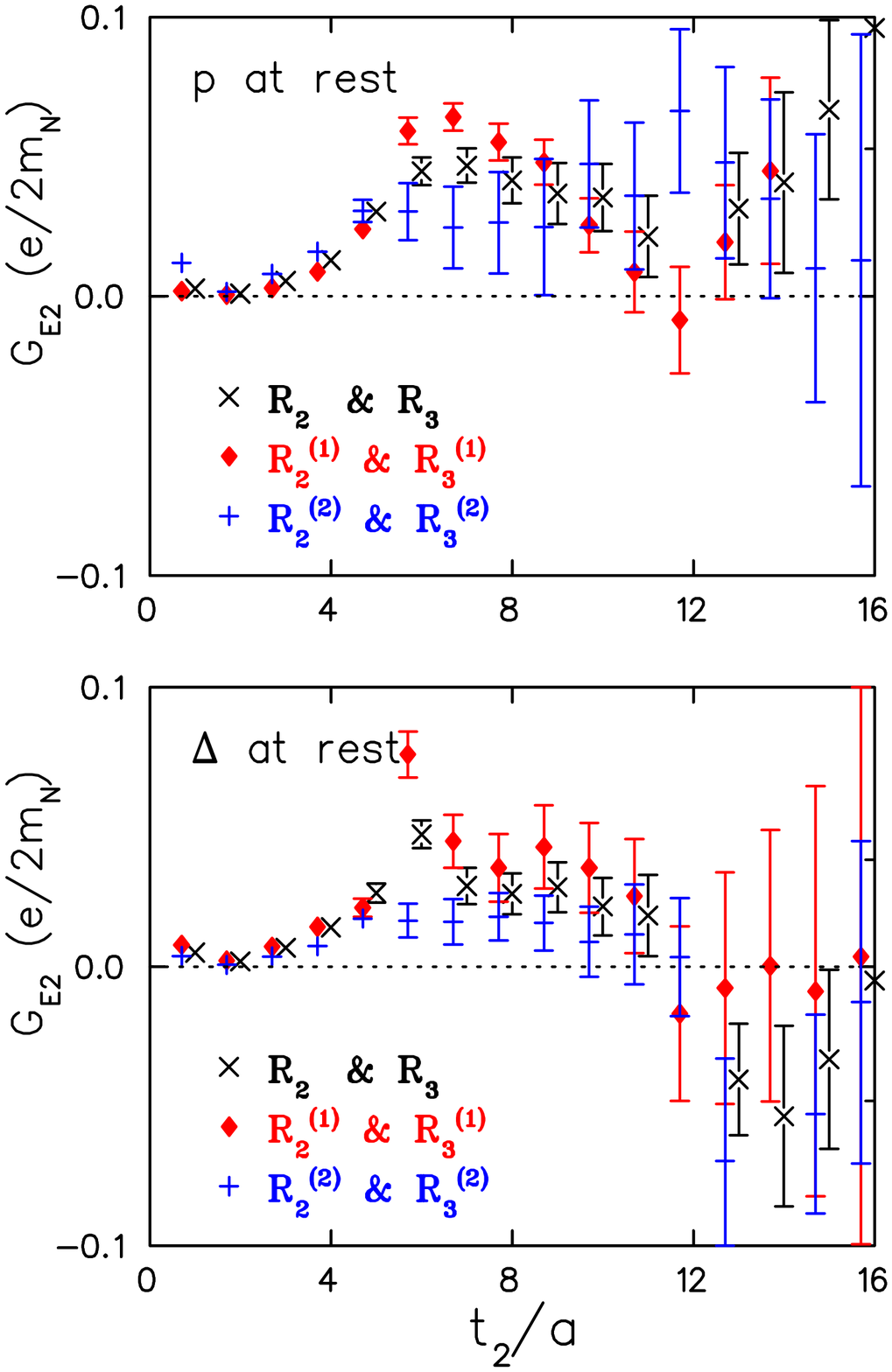}}
\caption{${\cal G}_{E2}^{(a)}$ in units of natural magnetons extracted
from Eq.~\ref{E2 24;31} 
 using the ratios $R_\sigma$ (x's),  $R_\sigma^{(1)}$ (diamonds) and  $R_\sigma^{(2)}$ (crosses)
 with the nucleon at rest (top)  and with  $\Delta$ at rest (bottom)
for dynamical quarks at $\kappa=0.1565$. 
The current couples to the quark at $t_1/a=6$. }
\label{fig:G vs R:GE2}
\end{figure}

\begin{figure}[h]
\epsfxsize=7.0truecm
\epsfysize=11truecm
\mbox{\epsfbox{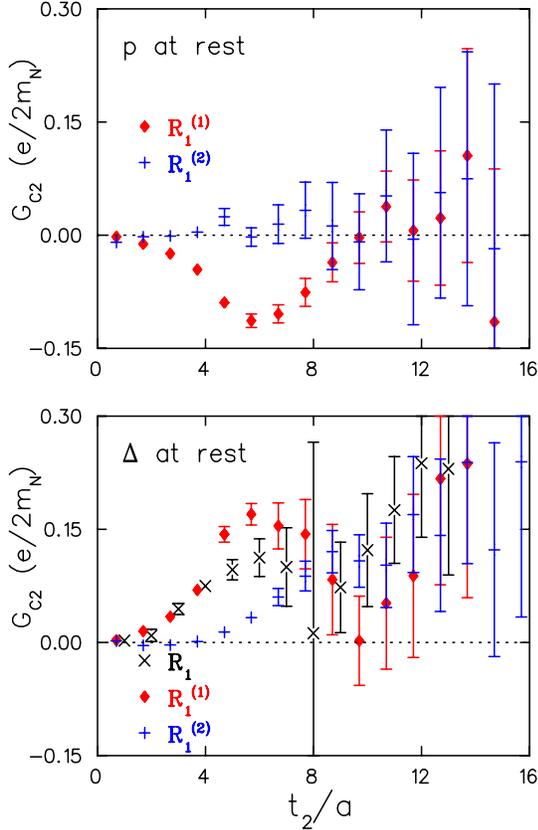}}
\caption{${\cal G}_{C2}^{(a)}$ in units of natural magnetons extracted
from Eq.~\ref{C2 11} 
 using the ratios $R_\sigma$ (x's),  $R_\sigma^{(1)}$ (diamonds)  and  $R_\sigma^{(2)}$ (crosses)
 with the nucleon at rest (top)  and with  $\Delta$ at rest (bottom)
for dynamical quarks at $\kappa=0.1565$. 
The current couples to the quark at $t_1/a=6$.}
\label{fig:G vs R:GC2}
\end{figure}

Smearing is essential 
for achieving   ground state dominance
before the signal from the time correlators 
is lost in the noisy large time limit.
We use the gauge invariant Wuppertal smearing,
$d(x,t) \rightarrow d^{\rm smear}(x,t)$,
 at the source and the sink. 
We smear  the fermion interpolating fields according to~\cite{Guesken,Alexandrou} 
\be
d^{\rm smear}({\bf x},t) = \sum_{\bf z} F({\bf x},{\bf z};U(t)) d({\bf z},t)
\ee
with the gauge invariant smearing function constructed from the
hopping matrix $H$: 
\be
F({\bf x},{\bf z};U(t)) = (1+\alpha H)^n({\bf x},{\bf z};U(t)),
\ee
where $H({\bf x},{\bf z};U(t))= \sum_{i=1}^3 \biggl( U_i({\bf x},t)\delta_{{\bf x,y}-i} +  U_i^\dagger({\bf x}-i,t)\delta_{{\bf x,y}+i}\biggr)$.
The parameters $\alpha=4.0$ and $n=50$ are 
tuned such as to optimize
the overlap with the baryon states.
 Quark
propagators with a photon insertion are computed with the sequential
source technique.
 Therefore for the three-point functions we require
two inversions with the second inversion having a momentum dependent source.
The sequential source technique requires that the photon couples to the
quark at fixed time $t_1$ which is chosen large enough so that the nucleon and
$\Delta$  ground states are identified.  For the lattices used here
we know from the nucleon and $\Delta$ two point functions that 
for $t_1\ge 5a$ the excited state contributions become negligible.  
We also make use of the equal weighting of 
$\lbrace U \rbrace $ and $\lbrace U^* \rbrace $ gauge configurations  
in the lattice action~\cite{Draper} and
the parity symmetry of our correlators to improve the
plateau behaviour of the reduced ratio 
$\Pi_\sigma({\bf p}', {\bf p}\; ; \Gamma ; \mu)$. 
For implementation of this
procedure we need an additional sequential propagator with the photon
 carrying momentum  $-{\bf q}$. To see
the improvement we compare in Fig.~\ref{fig:noisereduction} 
the results obtained when
for each configuration we consider the photon carrying 
 either momentum  ${\bf q}$ or   $-{\bf q}$ 
to those obtained for 
the photon with 
momenta ${\bf q}$ {\it and}  $-{\bf q}$ 
for equal statistics.
 As it can be seen, for large time separations, 
the quality of the  plateau obtained for the reduced ratio
$ \Pi_{\sigma}({\bf p}^{\; \prime}, {\bf p}\; ; \Gamma ; \mu)$
is improved 
when the  equal reweighting of  $\lbrace U \rbrace $ and 
$\lbrace U^* \rbrace $ is implemented enabling us to fit
over a larger range,
 far from the time insertion of the current.
 In all the results presented here this 
 benefit is utilized:  thus
for each configuration the sequential propagator is inverted 
twice,
once with the photon carrying momentum ${\bf q}$ and once
carrying momentum $-{\bf q}$. Unambiguous identification of the plateau
region at large time separations from the source 
outweighs the additional costs.

The ratios given in
Eqs.~\ref{R-ratio},~\ref{R1-ratio} and \ref{R2-ratio} provide
three ways for extracting the form factors.
In Fig.~\ref{fig:G vs R}
 we compare  these three possibilities for evaluating
${\cal G}_{M1}^{(a)}$.
The plot 
illustrates the results for the unquenched case at $\kappa=0.1565$ --
a similar behaviour is observed in all cases.
We consider two kinematically different cases:
 one where the nucleon has zero momentum and therefore the 
$\Delta$ carries momentum ${\bf q}$ and the other\
 where the $\Delta$ is produced 
at rest and therefore the nucleon has
momentum $-{\bf q}$. What  is clearly seen for both kinematics, is that  
$R_\sigma$ as given in Eq.~\ref{R-ratio} yields
 the best plateau,  which starts as early as two time slices
away from 
the time  where the  current couples to the quark. In contrast, 
the other two definitions require five time slices  to show
convergence to the same value. Evidently excited
states contributions come with the opposite sign
in the matrix elements $\gamma N \rightarrow \Delta$ and 
$ \Delta \rightarrow \gamma N$
canceling 
to a large extent in the ratio 
$R_\sigma$. The same conclusion
is reached for ${\cal G}_{M1}^{(b)}$  extracted from
Eq.~\ref{M1 31;13} (without the $m_\Delta/E_\Delta $ factor 
when $\Delta$ is at rest).
In Fig.~\ref{fig:G vs R:GE2} we show the analogous results for 
${\cal G}_{E2}^{(a)}$ extracted from Eq.~\ref{E2 24;31}.
Again the ratio $R_\sigma$  yields an earlier
plateau, which in this case is indispensable since
the signal becomes too noisy beyond time separations $t_1-t_2 \ge 6/a$. 
  Eq.~\ref{E2 31;13}  provides an alternative
way to extract the electric quadrupole form factor. However  the plateaus
for the reduced ratios involved in the extraction of  ${\cal G}_{E2}^{(b)}$
deteriorate when the nucleon is at rest leading to unreliable results.
When the $\Delta$ is at rest,   Eq.~\ref{E2 31;13} 
produces a good plateau  for the ratio $R_\sigma$  
yielding results for ${\cal G}_{E2}^{(b)}$
consistent with those for ${\cal G}_{E2}^{(a)}$.
Therefore in what follows,  Eq.~\ref{E2 31;13} will only be used when the
$\Delta$ is produced at rest to check consistency with 
the values obtained from  Eq.~\ref{E2 24;31}.
The ratio $R_{EM}$ is evaluated
using   ${\cal G}_{E2}^{(a)}$ and
${\cal G}_{M1}^{(a)}$ since this is applicable for both types of
kinematics.
Systematic errors and noise 
become worse for ${\cal G}_{C2}$. When the $\Delta$ is
produced in motion the ratio $R_\sigma$ cannot be used since the correlators
$G_\sigma^{\Delta j N}$ and $G_\sigma^{N j \Delta}$ do not agree even
in the sign for time separations $t_2-t_1 \le 11/a$ and after that
they are too noisy to be usable. This is seen in Fig.~\ref{fig:G vs R:GC2}
where we show results for ${\cal G}_{C2}^{(a)}$. The other two
combinations that yield ${\cal G}_{C2}^{(b)}$ 
show at best equally poor results.
When the $\Delta$ is produced at rest the signs
are consistent and the ratio $R_\sigma$, although noisy,
 can be used to look for a plateau.
In  what follows we will
attempt to fit 
over the plateau range only for the case when the $\Delta$
is at rest in order to provide an estimate on  ${\cal G}_{C2}$ wherever
possible.

\begin{figure}[h]
\epsfxsize=8.0truecm
\epsfysize=5.5truecm
\mbox{\epsfbox{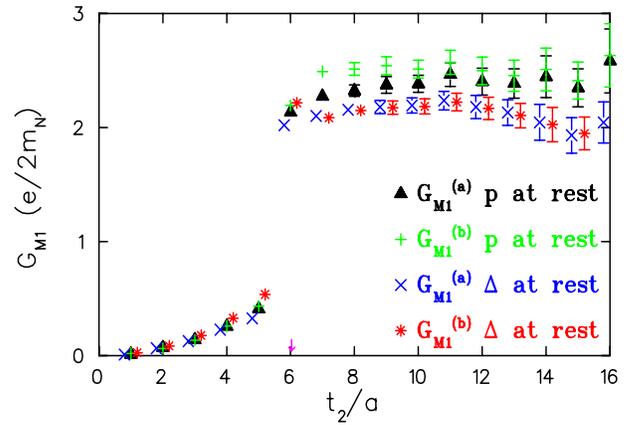}}
\caption{${\cal G}_{M1}$ in units of natural magnetons ($e/2m_N$) 
for the unquenched theory on a lattice of
size $16^3\times 32$ at $\kappa=0.1565$ and momentum
transfer ${\bf q}=(2\pi/16a,0,0)$.
Filled triangles
and crosses denote results for  ${\cal G}_{M1}^{(a)}$ (Eq.\ref{M1 24})
and  ${\cal G}_{M1}^{(b)}$ (Eq.\ref{M1 31;13}) 
respectively when  the nucleon at rest. The
x's and stars denote  respectively results  for  ${\cal G}_{M1}^{(a)}$ 
and  ${\cal G}_{M1}^{(b)}$ when the $\Delta$ is produced at rest.
They are shifted to the left and right
of the results obtained when the nucleon is at rest for clarity.  
The photon is injected at $t_1/a=6$
as shown by the arrow.} 
\label{fig:GM1 D0 & N0 sesam}
\end{figure} 

\begin{figure}[h]
\epsfxsize=8.0truecm
\epsfysize=5.5truecm
\mbox{\epsfbox{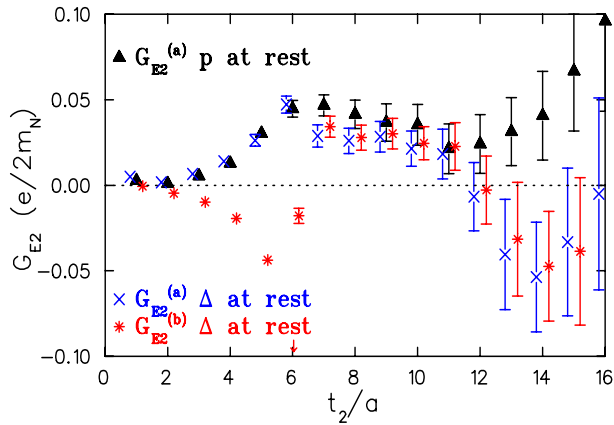}}
\caption{${\cal G}_{E2}$ in natural magnetons 
for the unquenched theory on a lattice of
size $16^3\times 32$ at $\kappa=0.1565$ and momentum
transfer ${\bf q}=(2\pi/16a,0,0)$. 
Filled triangles
 denote the results for ${\cal G}_{E2}^{(a)}$  when the nucleon is at rest.
The x's and stars denote respectively results for ${\cal G}_{E2}^{(a)}$  and 
${\cal G}_{E2}^{(b)}$ 
 when the $\Delta$ is produced at rest.
The rest of the notation is the same as that of Fig.~\ref{fig:GM1 D0 & N0 sesam}.
}
\label{fig:GE2 D0 & N0 sesam}
\end{figure}

\begin{figure}[h]
\epsfxsize=8.0truecm
\epsfysize=5.5truecm
\mbox{\epsfbox{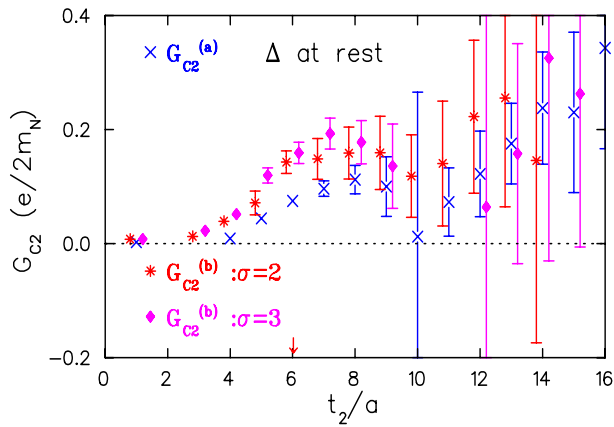}}
\caption{${\cal G}_{C2}$  in natural magnetons 
for the unquenched theory on a lattice of
size $16^3\times 32$ at $\kappa=0.1565$ and momentum
transfer ${\bf q}=(2\pi/16a,0,0)$. 
We show only results for the $\Delta$  at rest. The x's
denote the results for ${\cal G}_{C2}^{(a)}$  (without the factor 
 $m_\Delta/E_\Delta$ in Eq.~\ref{C2 11}), stars and diamonds denote results 
for ${\cal G}_{C2}^{(b)}$ using  Eq.~\ref{C2 22+33}
with the $\Delta$ vector index $\sigma=2$  
and  $\sigma=3$ respectively.
The rest of the notation is the same as that of Fig.~\ref{fig:GM1 D0 & N0 sesam}.}
\label{fig:GC2 D0 & N0 sesam}
\end{figure}

In the case of the dominant amplitude $M1$ definitions ${\cal G}_{M1}^{(a)}$  
and ${\cal G}_{M1}^{(b)}$ can be used for both our kinematics and 
the results 
for ${\cal G}_{M1}^{(a)}$   and ${\cal G}_{M1}^{(b)}$
are directly compared in Fig.~\ref{fig:GM1 D0 & N0 sesam} 
for the SESAM lattice at $\kappa=0.1565$
 at momentum transfer ${\bf q}=(2\pi/16a,0,0)$
for both  when the $\Delta$ is produced at rest and 
when  the nucleon is at rest.
In  the  $\Delta$  rest frame   ${\cal G}_{M1}^{(a)}$ 
and  ${\cal G}_{M1}^{(b)}$
are in perfect agreement 
with a plateau region that sets in as early as two time slices 
from the insertion of the current.
 When the $\Delta$  carries
momentum, ${\cal G}_{M1}^{(a)}$ and ${\cal G}_{M1}^{(b)}$
agree for time separations $t_2-t_1 \ge 4/a$ 
suggesting that contamination due to excited states is more severe
when the $\Delta$ is produced in motion.

Figs.~\ref{fig:GE2 D0 & N0 sesam} and \ref{fig:GC2 D0 & N0 sesam} show the 
corresponding results
for ${\cal G}_{E2}$ and ${\cal G}_{C2}$. 
For both form factors the data become very noisy for time  separations 
$t_2-t_1 \ge 7/a$. However for the case of the 
 electric quadrupole we have a plateau region 
extending over four time slices enabling us to extract a value
for ${\cal G}_{E2}^{(a)}$ both for $\Delta$ static and 
for $\Delta$ carrying a momentum.
When the $\Delta$ is produced at rest
the reduced ratio needed for the extraction of  ${\cal G}_{E2}^{(b)}$
shows similar plateau behaviour as that obtained for  ${\cal G}_{E2}^{(a)}$
 whereas, as we already
mentioned, when the $\Delta$ is not at rest this
correlator is too noisy to be useful especially
for the small quenched and unquenched lattices.
  The identification of
the plateau region becomes particularly difficult for  ${\cal G}_{C2}$.
In Fig.~\ref{fig:GC2 D0 & N0 sesam} we show results
for the ratio $R_\sigma$ only for the case
when the $\Delta$ is produced at rest since, as  we have already discussed, 
when the $\Delta$ has non-zero momentum the three
point functions fluctuate in sign making the ratio $R_\sigma$ unusable.
From Fig.~\ref{fig:GC2 D0 & N0 sesam}
it can be seen  that  the results obtained for  ${\cal G}_{C2}^{(a)}$,
involving matrix elements with the $\Delta$  vector index $\sigma$
 in the same direction
as that of the momentum transfer,   have 
overall 
smaller statistical errors as compared to those for  ${\cal G}_{C2}^{(b)}$.
In fact, for most cases, the ratio $R_\sigma$ from Eqs.~\ref{C2 22+33}
can only be 
 determined
for small time separations, $t_2-t_1$,
insufficient to  
neglect contributions from excited states.
Therefore in what follows
 we will mostly use Eq.~\ref{C2 11} to estimate  ${\cal G}_{C2}$.

The behaviour of the transition form factors 
shown in Figs.~\ref{fig:GM1 D0 & N0 sesam},~\ref{fig:GE2 D0 & N0 sesam} 
and \ref{fig:GC2 D0 & N0 sesam} for the SESAM lattice at $\kappa=0.1565$ is
typical  and it is observed for the other lattices and $\kappa$ 
values.

\section{Results}
 The quenched calculation of the transition
form factors is carried out at  $\beta=6.0$ using lattices
of size $16^3\times 32$ and $32^3\times 64$. 
For the unquenched calculation we use the SESAM configurations 
at $\beta=5.6$ on a lattice of size $16^3\times 32$.
We use Wilson fermions with  hopping parameters $\kappa$
given in Table~\ref{table:parameters} where we also 
list
the values 
of the ratio  of the pion mass  to the rho mass.
For all configurations, the number of which is given in 
Table~\ref{table:parameters}, we double the statistics 
performing the calculation
both for ${\bf q}$ and for $-{\bf q}$.

\begin{table}
\caption{\label{table:parameters} 
 $\kappa$ values and  momentum transfers
 used for the evaluation of the transition
form factors. The ratio of the rho mass to the pion mass at these $\kappa$
values is also given. We used the nucleon mass to set the scale.}
\medskip
\begin{tabular}{|c|c|c|c|c|c|} \hline
\multicolumn{6}{|c|} {Quenched $\beta=6.0$ $16^3\times 32$ ${\bf q}^2$=0.64 GeV$^2$} \\ \hline  
 \multicolumn{2}{|c|}{$ Q^2$ (GeV$^2$)} &$\kappa$ & $m_\pi/m_\rho$ & $m_\pi$(GeV) & \# of confs \\ \hline
p at rest & $\Delta$ at rest & & & & \\  \hline
0.57 & 0.64   & 0.1530 & 0.84 & 0.877(3) &100 \\ 
0.55 & 0.64   & 0.1540 & 0.78 & 0.736(2) &  100 \\
0.50 & 0.64   & 0.1550 & 0.70 & 0.604(2) &100 \\ \hline
0.40 & 0.64   & $\kappa_c=0.1571$ & 0 & 0    & extrapolated \\ \hline
\multicolumn{6}{|c|} {Quenched $\beta=6.0$ $32^3\times 64$ $ {\bf q}^2=0.64$ GeV$^2$ } \\ \hline
 \multicolumn{2}{|c|}{$ Q^2$ (GeV$^2$)} &$\kappa$ & $m_\pi/m_\rho$ & $m_\pi$(GeV) & \# of confs \\ \hline
p at rest & $\Delta$ at rest & & &   &\\ \hline
0.50 & 0.64   & 0.1550 & 0.69 & 0.598(4) & 100 \\ \hline
\multicolumn{6}{|c|} {Quenched $\beta=6.0$ $32^3\times 64$ $ {\bf q}^2=0.16$ GeV$^2$ } \\ \hline
 \multicolumn{2}{|c|}{$ Q^2$ (GeV$^2$)} &$\kappa$ & $m_\pi/m_\rho$ & $m_\pi$(GeV) &\# of confs \\  \hline
p at rest & $\Delta$ at rest & & &  &\\ \hline
0.13 &  0.16  & 0.1554 & 0.64 & 0.537(4) & 100 \\  
0.10 &  0.15  & 0.1558 & 0.59 & 0.469(4) & 100  \\    
0.057 & 0.13  & 0.1562 & 0.50 & 0.392(4) &100   \\ \hline
0.064 & 0.13   & $\kappa_c=0.1571$ & 0 & 0   & extrapolated \\ \hline   
\multicolumn{6}{|c|} {Unquenched $\beta=5.6$ $16^3\times 32$ $ {\bf q}^2=0.54$ GeV$^2$  }\\ \hline
 \multicolumn{2}{|c|}{$ Q^2$ (GeV$^2$)} &$\kappa$ & $m_\pi/m_\rho$ &$m_\pi$(GeV) & \# of confs \\  \hline
p at rest & $\Delta$ at rest & & &  &\\ \hline
0.48 & 0.54   & 0.1560  & 0.83 & 0.837(6) & 196\\  
0.48 & 0.54   & 0.1565  & 0.81 & 0.755(10)& 200  \\
0.45 & 0.54  & 0.1570  & 0.76 & 0.655(9) &201  \\
0.45 & 0.54  & 0.1575  & 0.68 & 0.527(8) & 200  \\ \hline
0.40 & 0.53   & $\kappa_c=0.1585$ & 0  & 0   & extrapolated \\ \hline   
\end{tabular}
\end{table}

\begin{figure}[h]
\epsfxsize=8.0truecm
\epsfysize=14.truecm
\mbox{\epsfbox{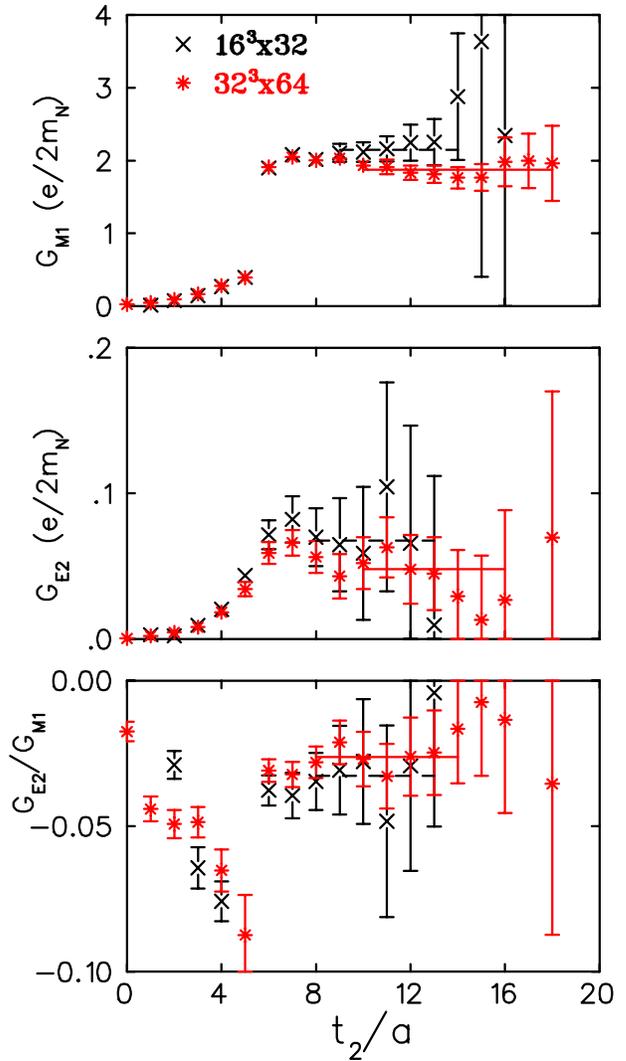}}
\caption{${\cal G}_{M1}^{(a)}$, ${\cal G}_{E2}^{(a)}$ in units of
 natural magnetons and 
$-{\cal G}_{E2}^{(a)}/{\cal G}_{M1}^{(a)}$ in the quenched theory
for lattices of size $16^3\times 32$ (x-symbols) 
and  $32^3\times 64$ (stars) at $\kappa=0.155$ 
and momentum transfer ${\bf q}=(2\pi/16a,0,0)$. 
For this comparison the data for
the large lattice are shifted by two time slices to the left
so that the coupling of the photon to the quark
occurs at the same time point on the graph.
The dashed and solid lines are the plateau values for the small  and
 larger lattices respectively.}
\label{fig:volume}
\end{figure}

\begin{widetext}

\begin{table}[h]
\caption{
 The plateau values of ${\cal G}_{M1}^{(a)}$, ${\cal G}_{E2}^{(a)}$
in units of $e/2m_N$ and $R_{EM}$ in \%
 for different  fit ranges are given together with 
the $\chi^2/{\rm d.o.f.}$ for the quenched and SESAM lattices 
in the rest frame of the $\Delta$.  
${\cal G}_{C2}^{(a)}$ is also given for the cases where a plateau
can be identified. With star we indicate the value that we adopted. 
}
\label{fit ranges} 
\begin{tabular}{|c|c|c|c||c|c|c|c||c|c|c|c|} 
\hline
\multicolumn{4}{|c||} {$16^3\times 32$ quenched} & 
\multicolumn{4}{|c||} {$16^3\times 32$ unquenched} & 
\multicolumn{4}{|c||} {$32^3\times 64$ quenched}  \\
\hline \hline
\multicolumn{4}{|c||} {$\kappa=0.153$} & 
\multicolumn{4}{|c||} {$\kappa=0.1560$} &
\multicolumn{4}{|c||} {$\kappa=0.1554$} \\
\hline \hline
  ${\cal G}_{M1}$ &  $t_i/a $ & $t_f/a$ &   $\chi^2/{\rm d.o.f.}$ &
  ${\cal G}_{M1}$ &  $t_i/a $ & $t_f/a$ &   $\chi^2/{\rm d.o.f.}$ &
  ${\cal G}_{M1}$ &  $t_i/a $ & $t_f/a$ &   $\chi^2/{\rm d.o.f.}$ \\ \hline
 2.25(6) &  8 & 12  & 0.1 &   2.35(5) &  8 & 12  & 0.2   &  2.98(6)  & 10 & 16 & 1.9 \\
 2.25(7)$^*$ & 8  & 16  & 0.2 &   2.36(5)$^*$  & 8  & 16  & 0.5   &  3.00(7)  & 10 & 20 & 1.6 \\
 2.28(7) & 10 & 12  & 0.1 &   2.37(6) & 10 & 12  & 0.002 &  3.07(8)  & 12 & 16 & 0.4 \\
 2.26(9) & 10 & 16  & 0.2 &   2.39(7) & 10 & 16  & 0.9   &  3.09(9)$^*$  & 12 & 20 & 0.5 \\ 
\hline
  ${\cal G}_{E2}$ &  $t_i/a $& $t_f/a$ &   $\chi^2/{\rm d.o.f.}$ &
  ${\cal G}_{E2}$ &  $t_i/a $& $t_f/a$ &   $\chi^2/{\rm d.o.f.}$ &
  ${\cal G}_{E2}$ &  $t_i/a $& $t_f/a$ &   $\chi^2/{\rm d.o.f.}$  \\ \hline
 0.036(10) & 8  & 12 & 0.3 &  0.004(7) &  8  & 12 &  2   &  0.049(15)$^*$ & 10 & 16 & 0.5 \\
 0.038(11)$^*$ & 8  & 16 & 0.4 &  0.003(7) &  8  & 16 & 1.4  &   0.046(16) & 10 & 20 & 1.1 \\
 0.037(15) & 10 & 12 & 0.6 & -0.010(11)& 10  & 12 & 1.0  &  0.047(23) & 12 & 16 & 0.8 \\ 
 0.042(17) & 10 & 16 & 0.5 & -0.012(13)$^*$& 10  & 16 & 0.5 &  0.038(26) & 12 & 20 & 1.2 \\ 
\hline
  $ R_{EM}$       &  $t_i/a $& $t_f/a$ &   $\chi^2/{\rm d.o.f.}$ &
  $ R_{EM}$       &  $t_i/a $& $t_f/a$ &   $\chi^2/{\rm d.o.f.}$ &
  $ R_{EM}$       &  $t_i/a $& $t_f/a$ &   $\chi^2/{\rm d.o.f.}$ \\ \hline
 -1.6(5)  & 8  & 12 & 0.3   & -0.12(29)   & 8 & 12  & 2.0    &  -1.6(5)$^*$  & 10 & 16 & 0.6 \\
 -1.7(5)$^*$  & 8  & 16 & 0.4   & -0.13(32)   & 8 & 16  & 1.4    &  -1.6(5)  & 10 & 18 & 1.0\\
 -1.8(6)  & 9  & 16 & 0.4 &  0.01(41) & 9 & 16 & 1.0    &  -1.5(5)  & 10 & 20 & 1.2 \\
 -1.6(6)  & 10 & 12 & 0.6   &  0.44(48)   & 10 & 12 & 1.0   &  -1.5(8)  & 12 & 16 & 0.8 \\
 -1.8(7)  & 10 & 16 & 0.5   &  0.52(56)$^*$   & 10 & 16 & 0.5   &  -1.2(8)  & 12 & 20 & 1.3 \\
\hline
   ${\cal G}_{C2} $ &  $t_i/a $& $t_f/a$ &   $\chi^2/{\rm d.o.f.}$ &
   ${\cal G}_{C2} $ &  $t_i/a $& $t_f/a$ &   $\chi^2/{\rm d.o.f.}$ &
   ${\cal G}_{C2} $ &  $t_i/a $& $t_f$ &   $\chi^2/{\rm d.o.f.}$ \\ \hline
  & & & &  &  &  &    & 0.18(7)   & 10 & 16 & 0.3\\
  & & & &  &  &  &    & 0.12(13)$^*$  & 12 & 16 & 0.1\\
\hline \hline
 \multicolumn{4}{|c||} {$\kappa=0.155$} &
 \multicolumn{4}{|c||} {$\kappa=0.1570$} &
 \multicolumn{4}{|c|} {$\kappa=0.1562$} \\
\hline\hline
${\cal G}_{M1}$ &  $t_i/a $& $t_f/a$ &   $\chi^2/{\rm d.o.f.}$ &
${\cal G}_{M1}$ &  $t_i/a $& $t_f/a$ &   $\chi^2/{\rm d.o.f.}$ &
${\cal G}_{M1}$ &  $t_i/a $& $t_f/a$ &   $\chi^2/{\rm d.o.f.}$ \\ \hline
 1.94(8)   & 8  & 12 & 0.4   &   1.92(5)   & 8  & 12  & 0.2  & 2.71(9)   & 10 & 16 & 1.1 \\
 1.92(9)   & 8  & 16 & 0.9   &   1.91(7)$^*$   & 8  & 16  & 0.1  & 2.72(9)   & 10 & 20 & 1.2\\
 1.95(10)$^*$  & 9  & 15 & 0.8 & 1.93(8) & 9 & 16 &  0.3    & 2.82(13)  & 12 & 16 & 0.2\\
 2.01(12)   & 10 & 12 & 0.02  &   1.95(9)   & 10 & 12 & 0.2   & 2.83(13)$^*$  & 12 & 18 & 0.3 \\
 1.94(13)  & 10 & 16 & 1.1   &   1.94(10)  & 10 & 16 & 0.1   & 2.84(14)  & 12 & 20 & 0.6 \\
\hline
${\cal G}_{E2}$ &  $t_i/a $& $t_f/a$ &   $\chi^2/{\rm d.o.f.}$ &
${\cal G}_{E2}$ &  $t_i/a $& $t_f/a$ &   $\chi^2/{\rm d.o.f.}$ &
${\cal G}_{E2}$ &  $t_i/a $& $t_f/a$ &   $\chi^2/{\rm d.o.f.}$ \\ \hline
 0.074(24) & 8 & 12 &  0.3  & 0.022(9 )   & 8  & 12 & 0.2  & 0.077(24)$^*$  & 10 & 16 & 0.5 \\
 0.072(25) & 8 & 16 &  0.3  & 0.023(10)   & 8  & 16 & 0.4  & 0.075(25)  & 10 & 20 & 0.8\\
 0.078(33)$^*$ & 9 & 15 &  0.3  & 0.023(13)$^*$  & 9   & 15 & 0.2  & 0.105(47) &  12  &16 & 0.2    \\
 0.077(37) & 10 & 12 & 0.2  & 0.027(14)   & 10 & 12 & 0.1  & 0.097(50) & 12 & 18 & 0.6 \\
 0.071(38) & 10 & 16 & 0.3  & 0.029(16)   & 10 & 16 & 0.4  & 0.093(50) & 12 & 20 &0.8\\ 
\hline
$R_{EM}$       &  $t_i/a $& $t_f$/a &   $\chi^2/{\rm d.o.f.}$ &
$R_{EM}$       &  $t_i/a $& $t_f$/a &   $\chi^2/{\rm d.o.f.}$ &
$R_{EM}$       &  $t_i/a $& $t_f$/a &   $\chi^2/{\rm d.o.f.}$ \\ \hline
 -3.9(1.3)    & 8  & 12 & 0.2   & -1.1(5)       & 8  & 12 & 0.2  & -2.9(9)$^*$  &10  & 16 & 0.4 \\
 -3.8(1.3)$^*$    & 8  & 16 & 0.2   & -1.2(5)  & 8  & 16 & 0.4  & -2.8(10)  &10  & 20 & 0.8 \\
 -4.0(1.7) & 9  & 16 & 0.2   &  -1.2(7)$^*$   & 9 & 16  &  0.4    & -3.7(1.7)& 12 & 16 & 0.3 \\
 -3.8(1.9)    & 10 & 12 & 0.2   & -1.4(7)       & 10 & 12 & 0.2  & -3.4(1.7)&12 & 18  & 0.6\\
 -3.6(1.9)    & 10 & 16 & 0.2   & -1.5(9)       & 10 & 16 & 0.4  & -3.2(1.8)& 12 & 20 & 0.9\\ 
\hline
   ${\cal G}_{C2} $ &  $t_i/a $& $t_f/a$ &   $\chi^2/{\rm d.o.f.}$ &
  ${\cal G}_{C2} $ &  $t_i/a $& $t_f/a$ &   $\chi^2/{\rm d.o.f.}$ &
  ${\cal G}_{C2} $ &  $t_i/a $& $t_f/a$ &   $\chi^2/{\rm d.o.f.}$ \\ \hline
 & & & & 0.11(3) & 8 & 12 & 0.2  &  0.25(9)  & 10 & 15 & 0.2\\
 & & & & 0.11(3)$*$ & 8 & 15 & 0.3  &   &  &  &  \\
 & & & & 0.12(5) & 10& 12 & 0.4  & 0.17(21)$^*$  & 12 & 15 & 0.1\\
 & & & & 0.13(5) & 10& 15 & 0.3  &   &  &  &  \\
\hline
\end{tabular}
\end{table}

\end{widetext}

 To set the lattice
spacing $a$ in the quenched theory one can use the well known value 
of the string tension. 
However since we want
to compare quenched  and unquenched results 
we need a determination which 
is applicable in both cases.
Since we are calculating matrix elements 
in
the baryon sector it is more appropriate  to
use the value extracted from the nucleon  mass in the chiral 
limit to set the scale.
In the quenched  case the value extracted
using the nucleon mass is $a^{-1}=2.04(2)$~GeV ($a=0.089$~fm) and 
in the unquenched
$a^{-1}=1.88(7)$~GeV ($a=0.106$~fm)
with a systematic error of about 15\%~\cite{SESAM} 
coming from the choice of chiral extrapolation ansatz. 
We note that if one uses
the rho mass at the chiral limit one obtains $a^{-1}$=2.3~GeV($a=0.087$ fm)
for both the quenched and the unquenched theory~\cite{QCDPAX,Gupta,SESAM},
 with a systematic error of about 10\%
due to the choice of fitting range and chiral extrapolation 
ansatz~\cite{QCDPAX}. Therefore using the nucleon  or the rho mass
to set the scale 
yields results for $a$  within the systematic uncertainties. 
We would like to stress here that in the ratios $R_{EM}$ and $R_{SM}$,
which are the experimentally measured quantities, the 
 lattice spacing does not explicitly enter and therefore
 a precise determination of a
is less crucial here than in other studies. 
In Table~\ref{table:parameters} we use the value for $a$ 
obtained from the nucleon
mass to convert the momentum transfer 
 ${\bf q}$  at which the  transition moments
 are evaluated to physical units.
The lattice four-momentum transfer, $Q^2$, 
depends on the $\kappa$ values 
through the nucleon and $\Delta$ 
 masses. 
These values are given  in Table~\ref{table:parameters}.
As already mentioned we consider two kinematically different cases,
one with  the $\Delta$ being  at rest and the other with the nucleon
at rest.  
For all $\kappa$ values, including $\kappa_c$,
 for the case when the $\Delta$ is produced at rest
the variation in $Q^2$ is negligible.
For the case when the nucleon is at rest 
the  variation  in $Q^2$ over the range of $\kappa$
values 
considered
 here is sizable, in particular
when ${\bf q}=(2\pi/32a,0,0)$ where  $Q^2$ changes by a factor of two.
We will come back to this point in section IV when we discuss
the chiral extrapolation of the results.
In all the evaluations we take $u$ and $d$ quarks of the same mass.
The current  couples to the quark at time slice $t_1 = 6a $, 
for the lattices of size $16^3\times 32$ and at 
 $t_1 = 8a $ for the lattice of size $32^3\times 64$.

\begin{figure}[h]
\epsfxsize=8.0truecm
\epsfysize=11truecm
\mbox{\epsfbox{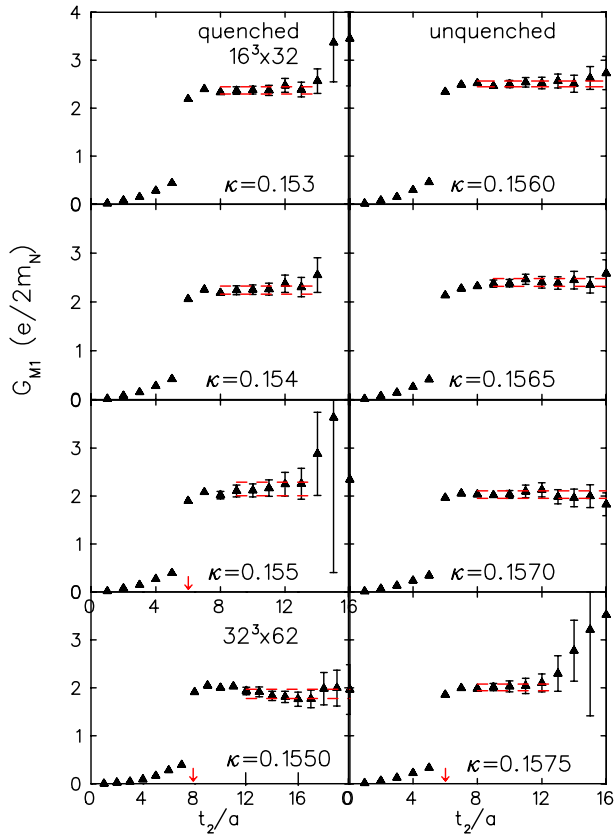}}
\caption{${\cal G}_{M1}^{(a)}$ at momentum transfer
 ${\bf q}=(2\pi/16a,0,0)$ in the rest frame of the nucleon. 
The three
upper  graphs on the left show the
quenched results for the lattice of size $16^3\time 32$
 at $\kappa=0.153,\>0.154$ and 0.155. The lowest graph on the left 
shows quenched results for the  lattice of size $32^3\time 64$.
On the right we show the unquenched results 
at  $\kappa=0.1560,\>0.1565,\>0.1570$ and 0.1575.
The photon couples to the quark 
at $t_1/a=6$ for the small
lattices and at $t_1/a=8$ for the large lattice
 as  shown by the arrow. The dashed
lines show the fit range and
bounds on the  plateau value obtained by jackknife analysis.}
\label{fig:GM1}
\end{figure}

\begin{figure}
\epsfxsize=6.5truecm
\epsfysize=9.truecm
\mbox{\epsfbox{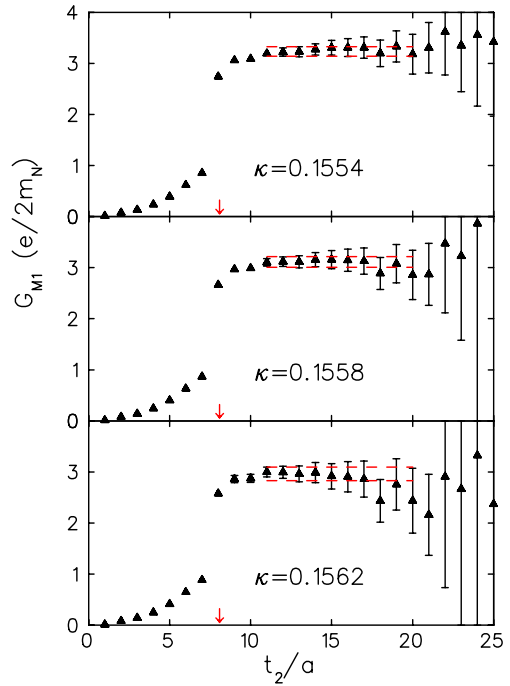}}
\caption{${\cal G}_{M1}^{(a)}$  for the quenched theory
on a lattice of size $32^3\times 64$ at  $\kappa=0.1554,\>0.1558$ and
0.1562  with the nucleon at rest and
for momentum transfer  ${\bf q}=2\pi/32a$.
The photon couples to the quark  at $t_1/a=8$ as indicated by the
arrow. The dashed
lines show the fit range and 
 bounds on the  plateau value obtained by jackknife analysis.}
\label{fig:GM1-quenched32}
\end{figure}

\begin{figure}[h]
\epsfxsize=8.0truecm
\epsfysize=10.5truecm
\mbox{\epsfbox{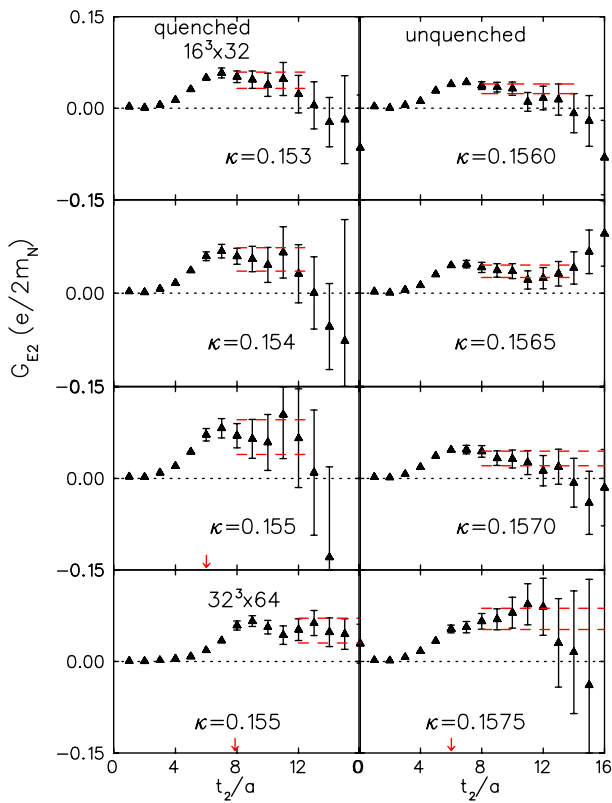}}
\caption{${\cal G}_{E2}^{(a)}$.
The notation is as in Fig.~\ref{fig:GM1}}.
\label{fig:GE2}
\end{figure}

\begin{figure}
\epsfxsize=7.0truecm
\epsfysize=9.truecm
\mbox{\epsfbox{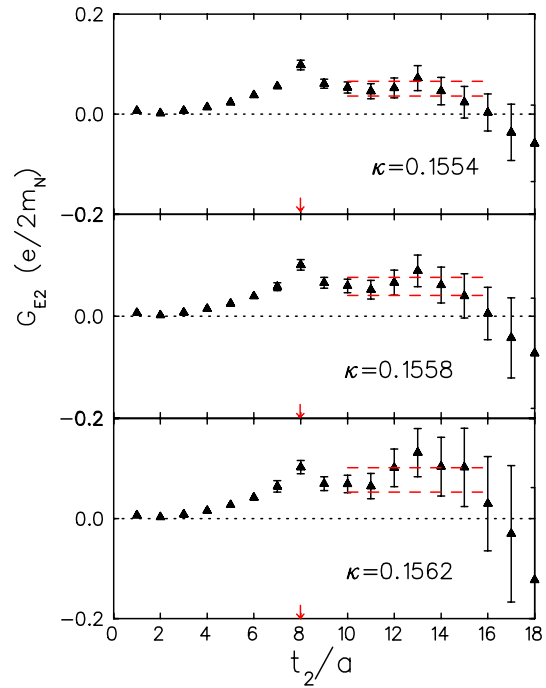}}
\caption{${\cal G}_{E2}^{(a)}$  for $\Delta$ at rest.
The notation is as in Fig.~\ref{fig:GM1-quenched32}.}
\label{fig:GE2-quenched32}
\end{figure}

\begin{figure}[h]
\epsfxsize=8truecm
\epsfysize=11truecm
\mbox{\epsfbox{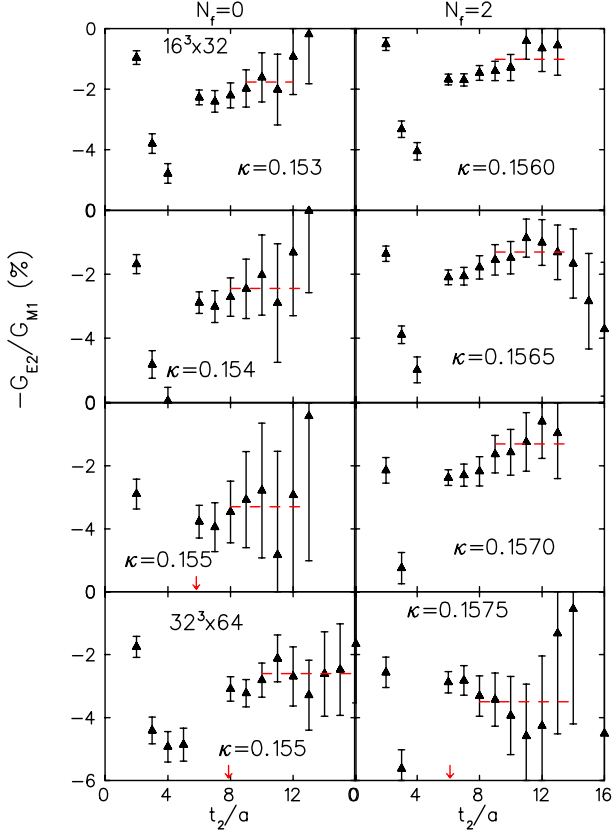}}
\caption{The ratio
$-{\cal G}_{E2}^{(a)}/{\cal G}_{M1}^{(a)}$ in the lab frame of 
the nucleon. 
The three upper graphs on the left show results for the quenched theory
on a lattice of size $16^3\times 32$ for ${\bf q}=(2\pi/16a,0,0)$
at $\kappa=0.153, \> 0.154$ 
0.155 and the lowest shows results
 for a lattice of size $32^3\times 16$ for the same value of ${\bf q}$
at $\kappa=0.155$. On the right we show results 
 for  the SESAM configurations at  $\kappa=0.1560,\>0.1565,\>0.1570$ and 0.1575
for ${\bf q}=(2\pi/16a,0,0)$.
The current couples to the quark at time 
 $t_1/a=6$ for the lattices of temporal
size $32a$  and
at $t_1/a=8$ for the lattice of temporal size $64a$ as shown by the arrow.}
\label{fig:EMR}
\end{figure}

To extract a reliable value for the form factors we identify the best plateau
region having the largest possible range and ensure that, changing the fit 
range within this plateau region, produces  results
that remain within statistical errors of each other.
In Table~\ref{fit ranges}
we give the  variation of the mean values from changing the fit range of the
plateaus for two representative $\kappa$ values
for each lattice in the rest frame of the $\Delta$. 
The quality of the plateaus and best fit ranges
 for the case when the nucleon is at rest is 
displayed in Figs.~\ref{fig:GM1} and  \ref{fig:GM1-quenched32} for
 ${\cal G}_{M1}$ and in Fig.~\ref{fig:GE2} for  ${\cal G}_{E2}$.
In Fig.~\ref{fig:GE2-quenched32} we display  the plateaus for the
 large quenched lattice for ${\cal G}_{E2}$ in the rest frame of
the $\Delta$ in order to have one case of
direct comparison to the numbers chosen for the best plateau
 in Table~\ref{fit ranges}. In Fig.~\ref{fig:EMR} we show the plateaus
and best fit ranges for $R_{EM}$ in the rest frame of the nucleon.
 From the plateaus values given in Table~\ref{fit ranges} 
it can be seen that the dependence of the 
results on the fit ranges are, in all cases, well  within 
the statistical errors obtained from jackknife analysis.

In order to access finite volume effects we calculate the transition
form factors in the quenched theory on two lattices of size $16^3\times 32$ and
$32^3\times 64$ at  the same momentum transfer. 
We show the results for  ${\cal G}_{M1}$,  ${\cal G}_{E2}$ and
the ratio  ${\cal G}_{E2}/{\cal G}_{M1}$ in Fig.~\ref{fig:volume}.
The data for
 ${\cal G}_{M1}$ 
obtained  on  the large volume lie systematically  below 
the ones obtained on the small lattice.
The results
for   ${\cal G}_{E2}$ as well as for the ratio
${\cal G}_{E2}/{\cal G}_{M1}$,
have overlapping errors. 
On the two volumes, the difference in the plateau values 
is smallest  in the ratio since volume effects
tend to cancel in numerator and denominator. 
The plateau values for  ${\cal G}_{M1}$ given in  Table~\ref{table:results}
are systematically lower  by 10-15\%  on the larger lattice.
Therefore 
there is a volume correction on the results obtained on the small lattices
of the order of 10\% which is comparable to the statistical error.
The results obtained on the larger lattice
have negligible volume dependence if  the 
volume correction scales inversely proportional to the volume.
 For ${\cal G}_{E2}$ and the ratio ${\cal G}_{E}/{\cal G}_{M1}$
which carry larger statistical errors  the volume dependence 
is harder to access since the difference in the values for the two volumes
is well within the statistical errors.

\begin{widetext}

\begin{table}
\caption{Results for ${\cal G}_{M1}$, ${\cal G}_{E2}$ 
in units of natural magnetons ($e/2m_N$) and for the
ratio  $-{\cal G}_{E2}/{\cal G}_{M1}$. } 
\begin{tabular}{|c|c|c|c|c|c|c|c|c|c|c|} 
\hline
\multicolumn{11}{|c|}{Quenched: $\beta=6.0$, $16^3\times 32$, ${\bf q}=(2\pi/16a,0,0)$ }  \\ \hline
$\kappa $  & \multicolumn{2}{|c|}{p at rest} & 
 \multicolumn{2}{|c|}{$\Delta$ at rest} & p at rest & 
 \multicolumn{2}{|c|}{$\Delta$ at rest} & p at rest & $\Delta$ at rest &
{$\Delta$ at rest} \\ \hline
  & ${\cal G}_{M1}^{(a)}$  & ${\cal G}_{M1}^{(b)}$  & 
 ${\cal G}_{M1}^{(a)}$  & ${\cal G}_{M1}^{(b)}$ &
 ${\cal G}_{E2}^{(a)}$  & ${\cal G}_{E2}^{(a)}$  & ${\cal G}_{E2}^{(b)}$   &
 \multicolumn{2}{|c|}{$R_{EM}$ (\%)} & 
${\cal G}_{C2}^{(a)}$    \\ \hline
0.153 &   2.37(7) & 2.52(7) &2.25(7) &2.26(7) & 0.046(13) & 0.038(11) & 0.035(12) & -1.8(8)  & -1.7(5)  &  \\
0.154 &   2.24(8) & 2.41(9) & 2.12(8)& 2.14(8)& 0.054(19) & 0.054(19) &0.045(18) & -2.4(9)  & -2.3(7)  &  \\
0.155 &   2.15(14)& 2.27(11)& 1.95(10)&1.99(10) & 0.068(28) & 0.078(33) &0.071(24) & -3.3(1.4)& -3.8(1.3)&  \\ \hline
\multicolumn{11}{|c|}{Quenched: $\beta=6.0$, $32^3\times 64$, ${\bf q}=(2\pi/16a,0,0)$ }  \\ \hline
$\kappa $  & \multicolumn{2}{|c|}{p at rest} & 
 \multicolumn{2}{|c|}{$\Delta$ at rest} & p at rest & 
 \multicolumn{2}{|c|}{$\Delta$ at rest} & p at rest & $\Delta$ at rest &
 {$\Delta$ at rest} \\ \hline
  & ${\cal G}_{M1}^{(a)}$  & ${\cal G}_{M1}^{(b)}$  & 
 ${\cal G}_{M1}^{(a)}$  & ${\cal G}_{M1}^{(b)}$ &
 ${\cal G}_{E2}^{(a)}$  & ${\cal G}_{E2}^{(a)}$  & ${\cal G}_{E2}^{(b)}$   &
 \multicolumn{2}{|c|}{$R_{EM}$ (\%)} & 
${\cal G}_{C2}^{(a)}$    \\ \hline
0.1550 &  1.87(10) & 1.99(10) & 1.78(7) & 1.76(7) & 0.048(20) & 0.049(40) &0.056(41)& -2.6(7) & -0.95(48) & \\ \hline
\multicolumn{11}{|c|}{Quenched: $\beta=6.0$, $32^3\times 64$, ${\bf q}=(2\pi/32a,0,0)$ }  \\ \hline
$\kappa $  & \multicolumn{2}{|c|}{p at rest} & 
 \multicolumn{2}{|c|}{$\Delta$ at rest} & p at rest & 
 \multicolumn{2}{|c|}{$\Delta$ at rest} & p at rest & $\Delta$ at rest &
 {$\Delta$ at rest} \\ \hline
  & ${\cal G}_{M1}^{(a)}$  & ${\cal G}_{M1}^{(b)}$  & 
 ${\cal G}_{M1}^{(a)}$  & ${\cal G}_{M1}^{(b)}$ &
 ${\cal G}_{E2}^{(a)}$  & ${\cal G}_{E2}^{(a)}$  & ${\cal G}_{E2}^{(b)}$   &
 \multicolumn{2}{|c|}{$R_{EM}$ (\%)} & 
${\cal G}_{C2}^{(a)}$    \\ \hline
0.1554 &  3.24(10) & 3.44(10) & 3.09(9) & 3.12(9) & 0.073(30) &0.049(15) & 0.049(16)& -2.2(9)      & -1.6(5) & 0.12(13)  \\
0.1558 &  3.11(11) & 3.35(10) & 2.97(11) & 3.00(11)& 0.079(23) & 0.059(18) & 0.057(18)  & -3.0(1.2)   & -2.1(6)   & 0.12(15)  \\
0.1562 &  2.96(13)& 3.22(12) & 2.83(13) & 2.84(14) & 0.131(53) & 0.077(24) & 0.075(24) & -4.1(2.0) & -2.9(9) & 0.17(21) \\ \hline 
\multicolumn{11}{|c|}{$N_f=2$: $\beta=5.6$, $16^3\times 32$  ${\bf q}=(2\pi/16a,0,0)$   } \\ \hline
$\kappa $  & \multicolumn{2}{|c|}{p at rest} & 
 \multicolumn{2}{|c|}{$\Delta$ at rest} & p at rest & 
 \multicolumn{2}{|c|}{$\Delta$ at rest} & p at rest & $\Delta$ at rest &
 {$\Delta$ at rest} \\ \hline
  & ${\cal G}_{M1}^{(a)}$  & ${\cal G}_{M1}^{(b)}$  & 
 ${\cal G}_{M1}^{(a)}$  & ${\cal G}_{M1}^{(b)}$ &
 ${\cal G}_{E2}^{(a)}$  & ${\cal G}_{E2}^{(a)}$  & ${\cal G}_{E2}^{(b)}$   &
 \multicolumn{2}{|c|}{$R_{EM}$ (\%)} & 
${\cal G}_{C2}^{(a)}$    \\ \hline
0.1560 & 2.51(6) & 2.65(6) & 2.36(5) &2.36(5) & 0.031(8)  & -0.012(13)& -0.011(12)&  -1.1(4) & 0.52(56)  &  -   \\
0.1565 & 2.34(8) & 2.52(7) & 2.13(6) & 2.15(6) & 0.035(10) & 0.020(11)& 0.025(9)& -1.3(5)  & -0.7(5)&0.11(3)  \\
0.1570 & 2.03(8) & 2.18(9) & 1.91(7) & 1.90(7) & 0.033(12) & 0.023(13)&0.027(10) & -1.3(7) & -1.2(7) & 0.11(3) \\ 
0.1575 & 2.01(7) & 2.29(8) & 1.73(6) & 1.72(7) & 0.070(17) & 0.021(16)&0.025(16) & -3.5(9)  & -1.2(9) & 0.08(3)  \\ 
 \hline
\end{tabular}
\label{table:results}
\end{table}

\end{widetext}

The values for ${\cal G}_{M1}$ extracted from the 
fits to the plateaus
are collected in Table~\ref{table:results}. The overall quality of the plateaus
as well as the best fit range chosen for each $\kappa$ value 
can be seen in Fig.~\ref{fig:GM1} 
for ${\bf q}=(2\pi/16a,0,0)$ and in Fig.~\ref{fig:GM1-quenched32}
for ${\bf q}=(2\pi/32a,0,0)$ for both the quenched and the unquenched theory.. 
The larger time extension of the lattice improves the identification
of the plateau region yielding reliable results for ${\cal G}_{M1}$
at the lighter quark masses.
From the values given in  Table~\ref{table:results}
 we conclude that, when the $\Delta$ is produced at rest, definitions 
${\cal G}_{M1}^{(a)}$ or  ${\cal G}_{M1}^{(b)}$ 
yield 
 the same values.
For the case where the $\Delta$ carries momentum ${\bf q}$ along
 the x-direction
${\cal G}_{M1}^{(b)}$, which is extracted from 
matrix elements for which the $\Delta$ vector index is  in 
the x-direction, yields consistently larger values than
those obtained from  ${\cal G}_{M1}^{(a)}$. This
difference in the values of  ${\cal G}_{M1}^{(a)} $ and ${\cal G}_{M1}^{(b)}$
is volume independent
and increases from about $6\%$ for the
heavier quarks to about $15\%$ for the lighter ones.
It is also independent of the value of the momentum carried by the $\Delta$.
It would be interesting to allow  
momentum transfers
 in all directions
to check if this difference will be reduced and also
use  an ${\cal O}(a)$ improved 
Dirac operator to check whether
it is due to finite 
$a$-effects.

To look for sea quark effects on the value of  ${\cal G}_{M1}$
we compare  the quenched results on the small lattice to
  the  unquenched results using the SESAM configurations 
at similar ratio of  the pion mass to the rho mass. From the
values of pion to rho mass ratios given in Table~\ref{table:parameters}  
one thus compares quenched results   at $\kappa=0.153,\;0.154$ and $0.155$ 
to the
unquenched results
 at $\kappa=0.1560,\; 0.1570$ and $0.1575$,
 respectively.
Unquenching leads to a stronger quark
mass dependence  increasing the value of ${\cal G}_{M1}$ at the
heaviest quark mass and reducing it at the two lighter quark masses. 

In Fig.~\ref{fig:GE2} we show the quenched and unquenched results for
${\cal G}_{E2}^{(a)}$  at ${\bf q}=(2\pi/16a,0,0)$ 
in the lab frame of the nucleon. 
The plateau region is limited to a a few time slices 
for the quenched lattice whereas for the SESAM lattice 
for twice the statistics the fits can be extended over a  larger
time interval.
For a static $\Delta$ both ${\cal G}_{E2}^{(a)}$
and ${\cal G}_{E2}^{(b)}$ show a similar 
behaviour and the best plateau fit ranges are given, for ${\cal G}_{E2}^{(a)}$,  in Table~\ref{fit ranges} for
representative $\kappa$ values.
As we already mentioned, although
the central value for   ${\cal G}_{E2}$ decreases
by going to the larger lattice,
the observed decrease is well within our statistics.
Similarly 
unquenching systematically  reduces the value of ${\cal G}_{E2}$. However 
with our statistics this reduction remains
within errors, for the three $\kappa$
values that correspond to similar ratios of pion to rho mass.
Plateau identification improves for the large lattice as  seen
in  Fig.~\ref{fig:GE2} where we display results at $\kappa=0.155$ for both the
large and small quenched lattice at equal momentum transfer.
This improvement is also seen in Fig.~\ref{fig:GE2-quenched32}, where we show results for
the large quenched lattice  
at ${\bf q}=(2\pi/32a,0,0)$, even though  lighter quarks
are used. Here part of the noise reduction
is due to having a smaller value of the momentum transfer. 
This enables to check the stability of our fits by changing the
fit range as given in Table~\ref{fit ranges}.

Our lattice results for the
 ratio $-{\cal G}_{E2}^{(a)}/{\cal G}_{M1}^{(a)}$
 are  displayed 
for the quenched and the unquenched theory in Fig.~\ref{fig:EMR}
in the rest frame 
of the nucleon. The general trend is that
this ratio  becomes more negative as
we approach the chiral limit.
The 
 plateau values seen in the figure and given in Table~\ref{table:results}
give no indication of an increase in this ratio as we unquench.
This may mean that pion cloud contributions, expected
to drive this ratio  
 more negative,  are suppressed at these large quark masses.
However one must keep in mind that part of the pion cloud is taken into
account in the quenched theory since, using relativistic quarks includes
backward propagation in time and thus effectively   
pionic contributions.
A study with lighter pions is required 
in order to assess 
the importance
of pion contributions due to sea quarks
on the value of  this ratio and thus on nucleon deformation.

\begin{figure}
\epsfxsize=7.0truecm
\epsfysize=9truecm
\mbox{\epsfbox{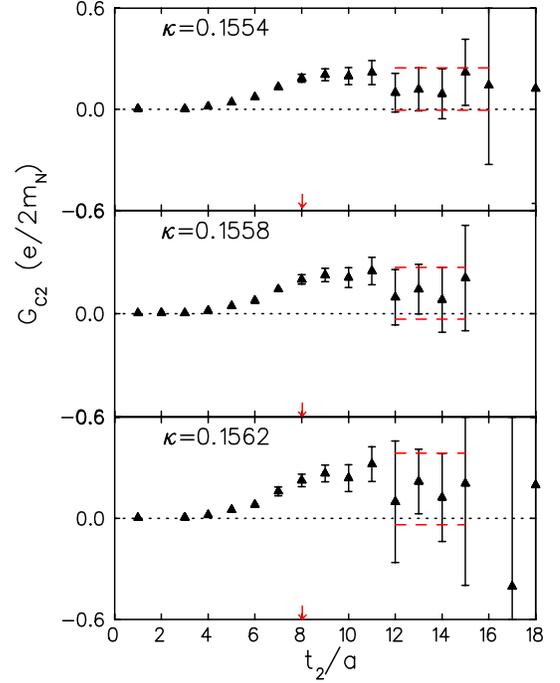}}
\caption{${\cal G}_{C2}^{(a)}$  in the rest  frame of the $\Delta$. 
The notation is as in Fig.~\ref{fig:GM1-quenched32}.}
\label{fig:GC2-quenched32}
\end{figure}

Finally in Fig.~\ref{fig:GC2-quenched32} we show the results for 
 ${\cal G}_{C2}^{(a)}$ with $\Delta$ static 
for the large quenched lattice for which 
 the ratio $R_\sigma$ can be obtained 
for long enough time
separations $t_2-t_1$ to allow plateau identification.
Although an evaluation of  ${\cal G}_{C2}^{(a)}$  is also carried out 
for the SESAM configurations a large
enough plateau range could not be identified for all $\kappa$ values.
Poor plateau identification leading to a  limited fit range
may introduce systematic errors  that are not included
in the errors of 
 the extracted values given in Table~\ref{table:results}.
However  for the cases where 
we can fit over 4 time slices far enough from the time of the current
insertion the mean value of ${\cal G}_{C2}^{(a)}$
is positive at all $\kappa$ values. 
Therefore even though the systematic errors are largest for this form factor,
which means that the jackknife errors given in Table~\ref{table:results}
provide an underestimation of the actual error, our lattice results
favour
a negative value   for the ratio
 $-{\cal G}_{C2}/{\cal G}_{M1}$ in agreement with experiment.

\section{Chiral extrapolations}

\begin{figure}
\epsfxsize=7.0truecm
\epsfysize=9.truecm
\mbox{\epsfbox{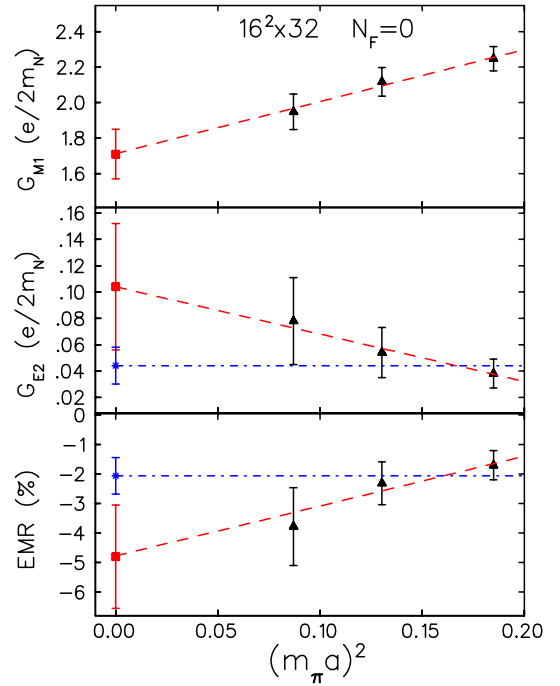}}
\caption{Top: ${\cal G}_{M1}^{(a)}$, middle: ${\cal G}_{E2}^{(a)}$ and 
bottom:EMR in \% versus $m_\pi^2$ in lattice units
 in the rest  frame of the $\Delta$ for the quenched $16^3\times 32$ lattice.
 The dashed line is  fit to $a+b m_\pi^2$ and the dashed-dotted to a constant
with the chiral value obtained shown with a filled square and star
for the two Ans\"atze respectively. }
\label{fig:chiral32}
\end{figure}

\begin{figure}
\epsfxsize=7.0truecm
\epsfysize=9.truecm
\mbox{\epsfbox{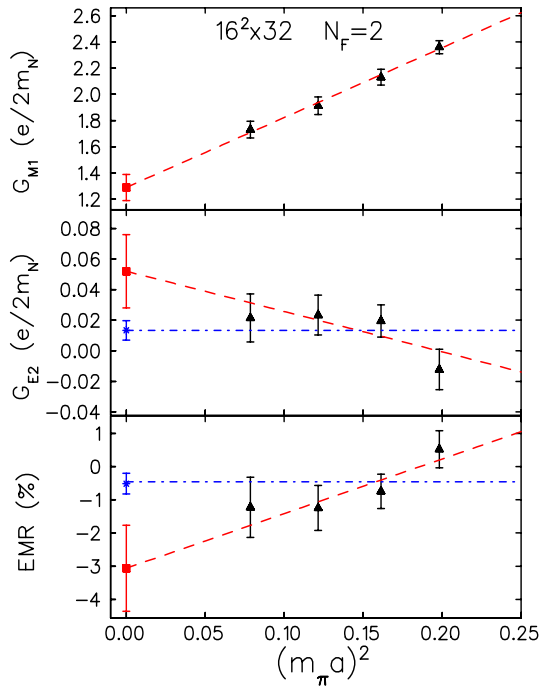}}
\caption{The same as in Fig.~\ref{fig:chiral32} but for the unquenched 
theory.}
\label{fig:chiralsesam}
\end{figure}

\begin{figure}
\epsfxsize=7.0truecm
\epsfysize=9.truecm
\mbox{\epsfbox{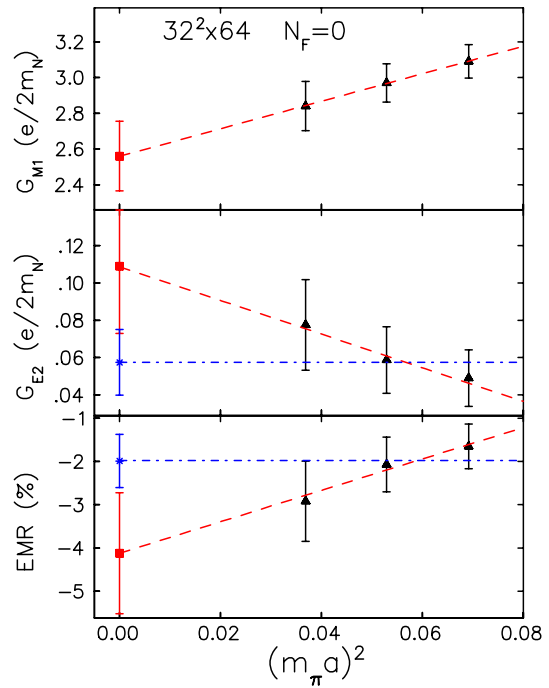}}
\caption{The same as in Fig.~\ref{fig:chiral32} but for the quenched 
$32^3\times 64$ lattice.}
\label{fig:chiral64}
\end{figure}

In order to obtain physical results we need to extrapolate the lattice data
to the chiral limit. 
Chiral perturbation theory has  been applied to calculate the
transition form factors~\cite{Gellas} but the range
of validity is limited to very small quark masses and
 very low momentum transfers and therefore it cannot
be used in the current analysis~\footnote{After the completion of this work chiral perturbation theory was also applied in the quenched case up to next to
 leading order by D. Arndt and B. C. Tiburzi, hep-lat/0308001. 
The main conclusion of this work is that
 ${\cal G}_{M1}$ and   ${\cal G}_{E2}$
depend logarithmically  on the pion mass squared
like in the full theory.}.
Since the nucleon or the $\Delta$ carry 
a finite momentum we expect chiral logs that appear at next-to-leading order
in chiral perturbation theory to be suppressed
for the momentum transfers studied in this work. 
Therefore a behaviour proportional to  
 the pion mass squared should be appropriate all the
way to very near the chiral limit. 
 The linear dependence of our results on the pion mass, $m_\pi$,
 squared is seen from 
Figs.~\ref{fig:chiral32}, \ref{fig:chiralsesam} and \ref{fig:chiral64}.
The results shown in these figures are for the
kinematical case where the $\Delta$ is stationary and  $Q^2$  remains 
almost unchanged as the pion mass approaches the chiral limit.
The errors on the lattice data are obtained by jackknife analysis and 
are purely statistical. Given the fact that we have results only at three
values of $\kappa$ in the quenched case and four in the unquenched case
 a linear fit is the best option we have.
For the quenched calculation the error on the extrapolated value at the chiral limit 
that we quote in Table~\ref{table:chiral} is obtained by 
doing a complete jackknife analysis for the three $\kappa$ values.
In the unquenched case, configurations at each 
quark mass are obtained from different Monte Carlo simulations, 
and a standard $\chi^2$-minimization procedure, which assumes 
uncorrelated  data points at each quark mass, can be applied.

\begin{figure}
\epsfxsize=7.0truecm
\epsfysize=9.truecm
\mbox{\epsfbox{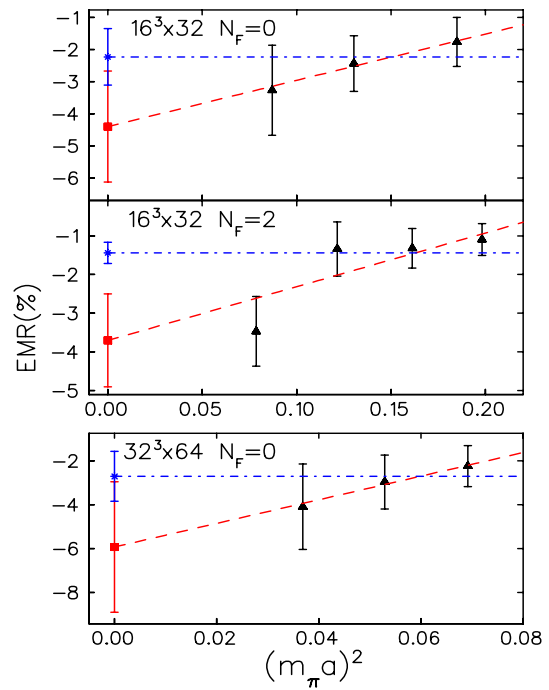}}
\caption{The EMR ratio in \% in the rest frame of the nucleon.
Top for the quenched $16^3 \times 32$ lattice,
middle for the unquenched theory and bottom for the quenched $32^3 \times 64$ lattice . 
The dashed line is  fit to $a+b m_\pi^2$ and the dashed-dotted to a constant
with the chiral value obtained shown with a filled square and star
for the two Ans\"atze respectively.}
\label{fig:chiral_EMR}
\end{figure}

Extrapolating linearly in  $m_\pi^2$
 we obtain for both quenched and unquenched lattices
 very good fits
for ${\cal G}_{M1}$, which is the most accurately determined quantity.
All the data points fall nicely on a straight line even in the unquenched
case where the data at different quark masses are uncorrelated  since they are 
obtained using a different set of configurations.
A linear Ansatz also fits well the quenched results for
${\cal G}_{E2}$ and $R_{EM}$ as can be seen in
Figs.~\ref{fig:chiral32}, \ref{fig:chiralsesam} and \ref{fig:chiral64}. 
For the corresponding unquenched results, although a linear
Ansatz still provides a good fit giving
a $\chi^2/{\rm d.o.f.}=1.1$ for ${\cal G}_{E2}$  and
$\chi^2/{\rm d.o.f.}=0.6$ for $R_{EM}$, the unquenched
results for ${\cal G}_{E2}$ and  $R_{EM}$ show very weak mass dependence
at the three lightest quark masses.
Chiral 
perturbation theory suggests
a similar  mass dependence for ${\cal G}_{E2}$ 
as that for  ${\cal G}_{M1}$. To check 
this mass dependence an unquenched calculation 
with higher statistics on a
 larger lattice to avoid finite size effects that, especially at the lightest
quark mass  ($\kappa=0.1575$), can be significant
 is called for.
Given this weak mass dependence we  also fit
 ${\cal G}_{E2}$ and  $R_{EM}$ to a constant which 
gives a lower  value
 for   ${\cal G}_{E2}$ and the absolute magnitude
of $R_{EM}$ at the chiral limit. For comparison we include
also for the quenched results the value obtained using a constant fit
even though a linear fit in $m_\pi^2$ is favored by the data. 
Although the chiral absolute value of  $R_{EM}$ is reduced by modifying
the fitting Ansatz to a constant,  $R_{EM}$ remains negative in all cases.

For kinematics 
where the nucleon is at rest $Q^2$
changes as we approach the chiral limit. 
This change is 
particularly
 severe for  the large
quenched lattice where
one would need the extrapolation of  
form factors computed at  $Q^2$ that decrease by 50\%.  
In phenomenological studies one models the $Q^2$ dependence of the three form 
factors by
\be
{\cal G}_{a}(Q^2) = {\cal G}_a(0) R_a (Q^2) G_E^p(Q^2)
\ee
where $R_a (Q^2)$ for $a=M1, E2$ and $C2$ measures the deviations 
from the proton electric form factor $ G_E^p(Q^2)=1/(1+Q^2/0.71)^2$. 
Usually experimental data are fitted by taking
$ R_{M1} (Q^2) =  R_{E2} (Q^2) =  R_{C2} (Q^2)  = 1 + \alpha \exp(-Q^2\gamma)$~\cite{Sato}.
There are not enough data to
simultaneously fit the $Q^2$ and the quark mass dependence. 
We are in the process of studying the $Q^2$ dependence
 of  these form factors
using the fixed source sequential technique~\cite{latt03,cairns03}. 
However, assuming
 that the above phenomenological Ans\"atze provide a good description 
for the $Q^2$ dependence of ${\cal G}_{M1}$ 
and ${\cal G}_{E2}$, we expect  the $Q^2$ dependence
to cancel  in the ratio $R_{EM}$. Therefore
a chiral extrapolation can be  performed for $R_{EM}$ 
also in the case of the nucleon
being at rest. A constant and a linear fit in $m_\pi^2$ are shown
for the three lattices in Fig.~\ref{fig:chiral_EMR}. Once more the constant
provides a lower limit for the absolute magnitude
of $R_{EM}$ and still leads to a negative value in the chiral limit 
in all cases.

\begin{figure}
\epsfxsize=7.0truecm
\epsfysize=10.truecm
\mbox{\epsfbox{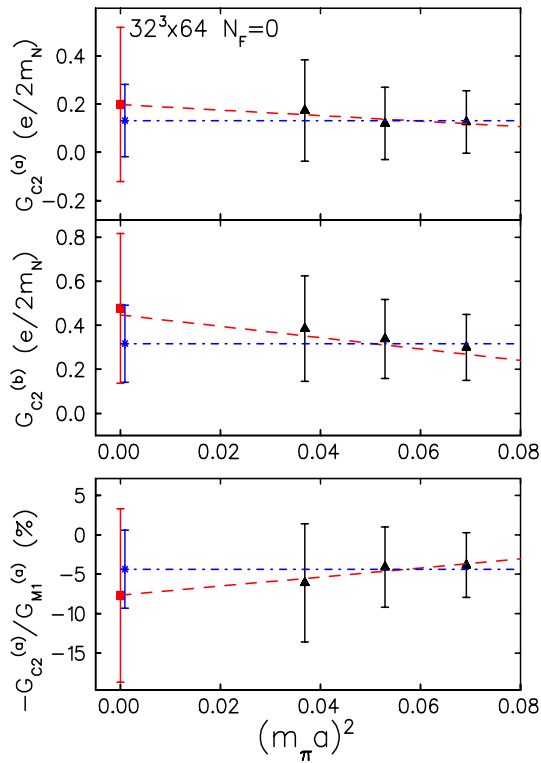}}
\caption{Top: ${\cal G}_{C2}^{(a)}$, middle: ${\cal G}_{C2}^{(b)}$ and 
bottom: $-{\cal G}_{C2}^{(a)}/ {\cal G}_{M1}^{(a)}$ 
in \% versus $m_\pi^2$ in lattice units
 in the rest  frame of the $\Delta$ for the quenched $32^3\times 64$ lattice.
 The dashed line is  fit to $a+b m_\pi^2$ and the dashed-dotted to a constant
with the chiral value obtained shown with a filled square and star
for the two Ans\"atze respectively.}
\label{fig:chiral_GC2}
\end{figure}

Finally we can not observe any  quark mass dependence for
${\cal G}_{C2}^{(a)}$ 
and ${\cal G}_{C2}^{(b)}$ as well as for the ratio 
$-{\cal G}_{C2}^{(a)}/{\cal G}_{M1}^{(a)} $  given the present 
statistical uncertainties
as can be seen
in Fig.~\ref{fig:chiral_GC2}. We included in the figure a constant 
and a linear fit yielding results that are consistent with  each other.
 We show only results obtained in the rest frame of the $\Delta$ 
for  the large quenched lattice where we have the best signal. 
For this lattice and kinematics  at $Q^2=0.13$~GeV$^2$ 
 the values we extract from the linear Ansatz are
 ${\cal G}_{C2}^{(a)}=  0.20(32) $ and  ${\cal G}_{C2}^{(b)}=  0.48(34) $ 
in units  of 
natural magnetons  $(e/2m_N)$. 
Changing
 the fitting range can lead to changes of the mean value of about 30\%
at each $\kappa$ value. Although the mean values of  ${\cal G}_{C2}^{(b)}$
are systematically larger than  those of ${\cal G}_{C2}^{(a)}$ at all $\kappa$ values they are always 
within statistical errors. Despite the  large
statistical and systematic uncertainties on this quantity the data favour
more a  negative than a positive
CMR ratio.

\section{Discussion} 
In Table~\ref{table:chiral} we list the chiral extrapolated values 
of the M1 and E2 form factors obtained in the rest frame of the $\Delta$
 as well as of the ratio $R_{EM}$ obtained for both kinematics. The values
given are the result of 
using a linear Ansatz in $m_\pi^2$ to extrapolate
to the chiral limit and the quoted errors are 
only statistical. As discussed in section III and IV  there is a
number of systematic errors that  we must consider when comparing the results 
of Table~\ref{table:chiral} to the experimental ones. 
We summarize here the source of systematic errors on the
form factors and their ratios. 
Finite volume effects are estimated by performing a quenched calculation
at the same value of momentum transfer on a lattice of size $16^3 \times 32$
and $32^3 \times 64$ at $\beta=6.0$. Assuming a 1/volume dependence for 
 ${\cal G}_{M1}$ the results show a 10-15\% correction for the data
obtained on the small lattice and negligible for the larger lattice.
The statistical errors on  ${\cal G}_{E2}$ are too large
to enable any volume correction to be
 extracted. However one would expect a similar volume dependence
as for ${\cal G}_{M1}$. On the other hand the volume dependence largely
cancels in the ratio and so we expect both EMR and CMR ratios to have
negligible volume dependence.
The variation in the mean values obtained by varying the plateau ranges
is smaller as compared to the statistical errors as can be seen from 
 Table~\ref{fit ranges}, where we have given the values obtained using
different fit ranges for representative $\kappa$ values for all the lattices
studied in this work. 
We look for unquenching effects by comparing quenched and unquenched
results on lattices of similar physical volume and lattice spacing
as well as  pion to rho mass ratio. For pion masses in the range
of 800-500 MeV unquenching effects are within statistical errors. 
 Extrapolation of 
the lattice results to the chiral limit represents the biggest
uncertainty that can only be eliminated by evaluating the form factors closer
to the chiral limit. 
Avoiding any chiral extrapolations
 the results on the two quenched and on the unquenched lattices at the lightest
quark mass give a ratio $R_{EM}$ in the range of about (-1 to -5)\%, which is
in accord with the range obtained in experimental measurements.
Both quenched and unquenched calculations are done at 
values of   $\beta$ corresponding to 
a lattice spacing of about 0.1~fm. Estimating finite lattice spacing effects
 would require repeating the
calculation at two larger values of $\beta$ keeping the physical volume
constant and attempting a continuum extrapolation. This is beyond 
our current computer resources and no systematic error due to the finite
lattice spacing can be estimated.

In the last column of Table~\ref{table:chiral} we give the results
for the ratio $R_{EM}$ obtained in the SL model 
using an effective Hamiltonian defined in the
subspace of $\pi N \oplus \gamma N \oplus \Delta$~\cite{Sato}.
 Within the SL model $R_{EM}$ is calculated in the rest frame
of the $\Delta$ as a function of
$Q^2$ for bare and dressed vertices. 
It is interesting that for bare vertices they obtain 
  non-zero values for $R_{EM}$ pointing
to the fact that a non-zero electric quadrupole amplitude implying
nucleon deformation is obtained even without pions.
This is in agreement with lattice
results on the rho deformation which is
observed in the quenched approximation
without pion contributions~\cite{wfs}. 
Comparing 
the value obtained in the SL model at $Q^2=0.64$~GeV$^2$ and $0.13$~GeV$^2$ using 
bare vertices
to our quenched lattice result
via   Eq.~\ref{EMR}~\cite{Sato} we see that 
quenched QCD produces a more negative value for $R_{EM}$ than  that
obtained in the SL model. 
One must however take into account that, in the
baryonic sector, quenching still 
includes part of the pion cloud due to quarks propagating backwards
in time. These are  absent in the SL model with bare vertices. On the other
hand 
unquenched lattice results at quark masses that correspond
to pion masses in the range 500-800~MeV do not influence
 the value of $R_{EM}$. This implies that
 pion cloud contributions from sea quarks at these pion masses are not large.
It remains an open issue whether this means that pionic contributions
from back-going quarks are also small, in which case
the difference between quenched lattice  and SL model results
cannot be explained solely by pion contributions from back-going quarks.
Results for $R_{EM}$ obtained in the SL model with fully dressed
vertices are twice as  negative
as those obtained with bare vertices 
showing, within this model,
 the importance of pionic contributions to deformation. 
Our unquenched results, obtained here for rather heavy pions,
 show no significant enhancement which may reflect 
that $q{\bar q}$ creation is suppressed. 
Since it is for light pions that large pionic contributions are expected
it is imperative to repeat the calculation
with dynamical quarks closer to the chiral limit to study sea quark contributions.

\begin{figure}
\epsfxsize=8.0truecm
\epsfysize=5.5truecm
\mbox{\epsfbox{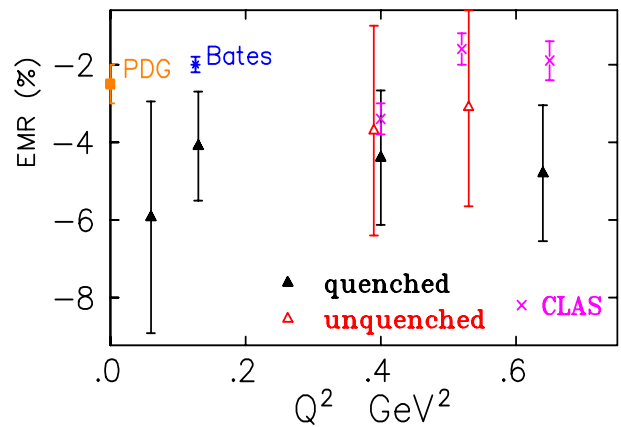}}
\caption{ The ratio
$R_{EM}$ in \%
versus $Q^2$ at the chiral limit. Filled triangles are quenched
and open triangles are unquenched results.
The filled square is the Particle Data Group result at $Q^2=0$~\cite{PDG},
the crosses are from ref.~\cite{Clas} and the star from ref.~\cite{Bates}.
The  errors shown on the experimental results are statistical.  
The lattice values shown  are those  obtained 
using a linear fit in $m_\pi^2$ to extrapolate to the chiral limit. 
In the unquenched case the error bars are increased so that 
an extrapolation to the chiral limit using a constant lies within the 
shown error bands.}
\label{fig:chiral}
\end{figure}

The values of the ratio  $R_{EM}$
given in Table~\ref{table:chiral} can be compared to those
measured at various values of $Q^2$
 in recent experimental searches for nucleon deformation.
The lattice results for $R_{EM}$ 
are plotted in Fig.~\ref{fig:chiral}
together with the recently measured
 experimental data. In the figure we also included
 the Particle Data group value for $R_{EM}$ 
at $Q^2=0$~\cite{PDG}. 
Once more what can be seen from this figure is that  quenched
and  unquenched results are within errors. 
The
unquenched results at our two available momentum transfers are in
agreement with the experimental results.
We stress that all the errors shown on this figure are statistical.
For the
experimental results the systematic error is about the same as
 the statistical error. For the lattice results
 the most
severe systematic error  comes from
assuming a linear Ansatz in  $m_\pi^2$ for the extrapolation
to the chiral limit. 

\begin{table}
\caption{
Results for ${\cal G}_{M1}$, ${\cal G}_{E2}$ 
in units of $e/2m_N$ and for the
ratio  $R_{EM}=-{\cal G}_{E2}/{\cal G}_{M1}$ in \% extrapolated to the chiral limit.
All errors given are statistical.
The last column gives the prediction for $ R_{EM}$ in \% 
obtained within the SL model~\cite{Sato}
where for the quenched case we quote their values 
without pion cloud
contributions 
and for the unquenched theory we give 
their fully dressed results 
for the case where the
$\Delta$ is produced at rest. In the last two columns
we give the  experimental results at similar values of $Q^2$.
} 
\medskip
\begin{tabular}{|c|c|c|c|c|c|} 
\hline
  &  $Q^2$~GeV$^2$ &  ${\cal G}_{M1}^{(a)}$ &  ${\cal G}_{E2}^{(a)}$ & 
\multicolumn{2}{|c|} { $ R_{EM}$ } \\ \hline
\multicolumn{5}{|c|}{Quenched QCD} & SL \\ \hline
p at rest        & 0.40 &    &   & -4.4(1.7)   &   \\
 $\Delta$ at rest & 0.64 &  1.71(14)  & 0.104(48) & -4.8(1.8) & -1.3\\
 $\Delta$ at rest & 0.13 &   2.56(20) & 0.108(36) & -4.1(1.4) & -1.3\\
 p at rest & 0.06 &    &  & -5.9(3.0)   &  \\ \hline
\multicolumn{5}{|c|}{Unquenched QCD} & SL \\ \hline
     $\Delta$ at rest       & 0.53 &  1.29(10)  & 0.052(24) & -3.1(1.3) & -2.7 \\ 
        p at rest   & 0.40 &   &  & -3.7(1.2)&   \\ 
\hline
\multicolumn{6}{|c|}{Experimental results}\\ \hline
  & 0.126 & & & -2.0(2)(2.0)~\cite{Bates} & \\
 & 0.40 &  &  & -3.4(4)(4)~\cite{Clas} & \\ 
 & 0.52  & & & -1.6(4)(4)~\cite{Clas}  & \\
\hline   
\end{tabular}
\label{table:chiral}
\end{table}  

From the lattice data obtained in this work 
$R_{SM}$ can only be estimated  in the chiral limit for the
large quenched lattice.  Extrapolating the results obtained in the $\Delta$ rest frame 
for the large quenched lattice
at $Q^2= 0.13$~GeV$^2$ we find
$-{\cal G}_{C2}^{(a)}/{\cal G}_{M1} =  (-7.7 \pm 11) \%$ which leads, via 
 Eq.~\ref{CMR}, to
$R_{SM}= (-1.2 \pm 1.7)$~\% where only the statistical error
is given. 
At $Q^2=0.126$~GeV$^2$ the experimental value of $R_{SM}$ 
is $(-6.5 \pm 0.2 \pm 2.5)\;\%$~\cite{Bates}.
 As already mentioned, the lattice evaluation of   this quantity
is affected by large systematic and statistical errors, which must be studied  before
 a more accurate determination can be obtained.

Finally, in order to facilitate comparison  with experiment and phenomenology,
not only for the ratios but also for  ${\cal G}_{M1}$ and  ${\cal G}_{E2}$
separately, 
we give the  relationship between
the Sachs form factors
studied in this work and the electromagnetic transition
amplitudes $f_{M1}$ and $f_{E2}$ in 
the rest frame of the $\Delta$~\cite{Sato}:
\be
f_{M1} = \frac{e}{2m_N} \biggl(\frac{|{\bf q}|m_\Delta}{m_N} \biggr) ^{1/2} 
\; \frac{{\cal G}_{M1}}{\biggl[1-q^2/(m_N+m_\Delta)^2\biggr]^{1/2}} 
\label{fM1}
\ee
\be
f_{E2} = -\frac{e}{2m_N} \bigl(\frac{|{\bf q}|m_\Delta}{m_N} \bigr) ^{1/2} 
\; \frac{{\cal G}_{E2}}{\bigl[1-q^2/(m_N+m_\Delta)^2\bigr]^{1/2}}
\label{fE2}
\ee
with $e=\sqrt{4\pi/137}$.
Since  these are  continuum relationships 
to  obtain results  for $f_{M1}$ and $f_{E2}$
from the values given 
in Table~\ref{table:chiral}
one  uses the physical nucleon and $\Delta$ mass.
We give the values for $f_{M1}$ and $f_{E2}$ in Table~\ref{table:experiment}.
They are
are related to the helicity amplitudes
$A_{1/2}$ and $A_{3/2}$ by
\beq
f_{M1}&=& -\frac{1}{2} (A_{1/2} + \sqrt{3} A_{3/2} ) \nonumber \\ 
f_{E2}&=& -\frac{1}{2} (A_{1/2} -  \frac{1}{\sqrt{3}}A_{3/2} ) \quad. 
\label{helicity}
\eeq
As can be seen  from Table~\ref{table:chiral},
 the values of the helicity amplitudes
for $Q^2$ closest to zero  are in agreement with the values at $Q^2=0$,
  $A_{1/2}=-0.135 \pm 0.006$~GeV$^{-1/2}$
and 
$A_{3/2}=-0.255 \pm 0.008$~GeV$^{-1/2}$, as
 given by the Particle 
Data Group.

\begin{table}
\caption{$f_{M1}$, $f_{E2}$, $A_{1/2}$ and $A_{3/2}$ in units of GeV$^{-1/2}$}
\begin{tabular}{|c|c|c|c|c|c|} 
\hline
 lattice &  $-q^2$~GeV$^2$ &  $f_{M1}$ &  $f_{E2}$ & $A_{1/2}$ &  $A_{3/2}$  \\ \hline
  $16^3\times 32$ & 0.64 &   0.266(22)  & -0.016(7) & -0.109(13)& -0.244(20) \\
 $32^3\times 64$ & 0.13 &   0.296(23) & -0.012(4)&-0.129(12) &  -0.267(20) \\
         SESAM   & 0.53 &  0.194(15)  & -0.0078(36) & -0.085(8) & -0.175(13)\\ 
\hline
\end{tabular}
\label{table:experiment}
\end{table}

\section{Conclusions}
The matrix element for the transition $\gamma N \rightarrow \Delta$
is computed in lattice QCD both for the quenched theory and for two dynamical
Wilson fermions in a  first attempt to study
unquenching effects on these form factors.
The dominant magnetic dipole
form factor is calculated with statistical accuracy of about  $10\%$.
The electric quadrupole is suppressed by an order of magnitude
and it is calculated to a statistical accuracy of about $50\%$.
Unquenching tends to decrease the values of ${\cal G}_{M1}$ and
${\cal G}_{E2}$.
The ratio of these form factors provides
a direct comparison to the experimentally measured ratio $R_{EM}$.
We find a negative value for   $R_{EM}$  of the order of a few \% 
in accord with experiment. For pions of
mass in the range of about $800-500$~MeV we obtain no evidence for
an increase in the value of $R_{EM}$  as  we 
unquench. It is expected
that pion cloud contributions are suppressed for these heavy
quarks and therefore it is important, in future studies, to use 
lighter dynamical quarks for the evaluation of these form factors.
Large statistical and systematic errors 
prevent at this stage
a determination of the Coulomb quadrupole  form factor.  
A detailed study of lattice artifacts 
will be needed  for better control of  systematic  errors,
before a more accurate
 extraction of  
the Coulomb form factor and of the ratio $R_{SM}$ 
is
possible.

{\bf Acknowledgments:} We thank H. Panagopoulos for providing a
Mathematica program for the $\gamma$-matrix algebra and 
D. Leinweber and T. Sato for discussions.

A.T. is supported by the Levendis Foundation and W.S. is partially supported by the Alexander von Humboldt Foundation . 
H. Neff acknowledges funding from 
 the European network ESOP (HPRN-CT-2000-00130) and  the
University of Cyprus. 

  This research used resources of the National Energy Research Scientific 
Computing Center, which is supported by the Office of Science of the U.S. 
Department of Energy under Contract No. DE-AC03-76SF00098. This work is 
supported in part by the U.S. Department of Energy (D.O.E.) under 
cooperative research agreement $\#$ DF-FC02-94ER40818 and
$\#$ DE-FC02-01ER41180.


\begin{thebibliography}{99}
\bibitem{error} 
The factor $m_\Delta/E_\Delta$ is missing in Eq.~3.25a 
and in the second term of Eqs.~3.26b and 3.27 of ref.~\cite{Leinweber}. 
The Eqs.~3.25a, 3.26b and 3.27 as given in  ref.~\cite{Leinweber} are
correct if the $\Delta$ is produced at rest.
\end{thebibliography}
\end{document}